\PassOptionsToPackage{unicode}{hyperref}
\PassOptionsToPackage{hyphens}{url}
\PassOptionsToPackage{dvipsnames,svgnames,x11names}{xcolor}
\documentclass[
  12pt]{article}

\usepackage{amsmath,amssymb}
\usepackage{iftex}
\ifPDFTeX
  \usepackage[T1]{fontenc}
  \usepackage[utf8]{inputenc}
  \usepackage{textcomp} 
\else 
  \usepackage{unicode-math}
  \defaultfontfeatures{Scale=MatchLowercase}
  \defaultfontfeatures[\rmfamily]{Ligatures=TeX,Scale=1}
\fi
\usepackage{lmodern}
\ifPDFTeX\else  
\fi
\IfFileExists{upquote.sty}{\usepackage{upquote}}{}
\IfFileExists{microtype.sty}{
  \usepackage[]{microtype}
  \UseMicrotypeSet[protrusion]{basicmath} 
}{}
\makeatletter
\@ifundefined{KOMAClassName}{
  \IfFileExists{parskip.sty}{%
    \usepackage{parskip}
  }{
    \setlength{\parindent}{0pt}
    \setlength{\parskip}{6pt plus 2pt minus 1pt}}
}{
  \KOMAoptions{parskip=half}}
\makeatother
\usepackage{xcolor}
\makeatletter
\ifx\paragraph\undefined\else
  \let\oldparagraph\paragraph
  \renewcommand{\paragraph}{
    \@ifstar
      \xxxParagraphStar
      \xxxParagraphNoStar
  }
  \newcommand{\xxxParagraphStar}[1]{\oldparagraph*{#1}\mbox{}}
  \newcommand{\xxxParagraphNoStar}[1]{\oldparagraph{#1}\mbox{}}
\fi
\ifx\subparagraph\undefined\else
  \let\oldsubparagraph\subparagraph
  \renewcommand{\subparagraph}{
    \@ifstar
      \xxxSubParagraphStar
      \xxxSubParagraphNoStar
  }
  \newcommand{\xxxSubParagraphStar}[1]{\oldsubparagraph*{#1}\mbox{}}
  \newcommand{\xxxSubParagraphNoStar}[1]{\oldsubparagraph{#1}\mbox{}}
\fi
\makeatother

\usepackage{longtable,booktabs,array}
\usepackage{calc} 
\usepackage{etoolbox}
\makeatletter
\patchcmd\longtable{\par}{\if@noskipsec\mbox{}\fi\par}{}{}
\makeatother
\IfFileExists{footnotehyper.sty}{\usepackage{footnotehyper}}{\usepackage{footnote}}
\makesavenoteenv{longtable}
\usepackage{graphicx}
\makeatletter
\def\maxwidth{\ifdim\Gin@nat@width>\linewidth\linewidth\else\Gin@nat@width\fi}
\def\maxheight{\ifdim\Gin@nat@height>\textheight\textheight\else\Gin@nat@height\fi}
\makeatother
\setkeys{Gin}{width=\maxwidth,height=\maxheight,keepaspectratio}
\makeatletter
\def\fps@figure{htbp}
\makeatother

\makeatletter
\@ifpackageloaded{caption}{}{\usepackage{caption}}
\AtBeginDocument{%
\ifdefined\contentsname
  \renewcommand*\contentsname{Table of contents}
\else
  \newcommand\contentsname{Table of contents}
\fi
\ifdefined\listfigurename
  \renewcommand*\listfigurename{List of Figures}
\else
  \newcommand\listfigurename{List of Figures}
\fi
\ifdefined\listtablename
  \renewcommand*\listtablename{List of Tables}
\else
  \newcommand\listtablename{List of Tables}
\fi
\ifdefined\figurename
  \renewcommand*\figurename{Figure}
\else
  \newcommand\figurename{Figure}
\fi
\ifdefined\tablename
  \renewcommand*\tablename{Table}
\else
  \newcommand\tablename{Table}
\fi
}
\@ifpackageloaded{float}{}{\usepackage{float}}
\floatstyle{ruled}
\@ifundefined{c@chapter}{\newfloat{codelisting}{h}{lop}}{\newfloat{codelisting}{h}{lop}[chapter]}
\floatname{codelisting}{Listing}

\makeatother
\makeatletter
\@ifpackageloaded{caption}{}{\usepackage{caption}}
\@ifpackageloaded{subcaption}{}{\usepackage{subcaption}}
\makeatother

\ifLuaTeX
  \usepackage{selnolig}  
\fi
\usepackage[]{natbib}
\usepackage{bookmark}

\IfFileExists{xurl.sty}{\usepackage{xurl}}{} 
\hypersetup{
  pdftitle={Title},
  pdfauthor={Author 1; Author 2},
  pdfkeywords={3 to 6 keywords, that do not appear in the title},
  colorlinks=true,
  linkcolor={blue},
  filecolor={Maroon},
  citecolor={Blue},
  urlcolor={Blue},
  pdfcreator={LaTeX via pandoc}}

\usepackage{multirow}
\usepackage{booktabs}
\usepackage{tikz}
\usetikzlibrary{arrows.meta}
\usetikzlibrary{arrows}
\usetikzlibrary{shapes.multipart}
\usetikzlibrary{positioning, fit}
\tikzset{every picture/.append style={scale=0.8, transform shape}}
\usepackage{enumitem}
\setlist[enumerate]{nosep, topsep=0pt, leftmargin=*}
\usepackage[a4paper, left=1.8cm, right=1.8cm, top=1.8cm, bottom=2.5cm]{geometry}

\newtheorem{assumption}{Assumption}
\newtheorem{definition}{Definition}[section]

\newtheorem{theorem}{Theorem}[section]
\newtheorem{proposition}{Proposition}[section]

\newtheorem{corollary}{Corollary}[section]

\makeatletter
\newcounter{algorithm}
\newcounter{breakablealgorithm}
\newenvironment{breakablealgorithm}
  {
   \begin{center}
     \refstepcounter{algorithm}
     \hrule width 1.0\linewidth height.8pt depth0pt \kern1pt
     \renewcommand{\caption}[2][\relax]{
          {\raggedright\textbf{\small Algorithm~\thealgorithm} ##2\par}%
       \ifx\relax##1\relax 
         \addcontentsline{loa}{algorithm}{\protect\numberline{\thealgorithm}##2}%
       \else 
         \addcontentsline{loa}{algorithm}{\protect\numberline{\thealgorithm}##1}%
       \fi
       \kern2pt\hrule width 1.0\linewidth \kern2pt
     }
  }{
     \kern2pt\hrule width 1.0\linewidth \relax
        \end{center}
  }
\makeatother

\newcommand{\cM}{{\mathcal{M}}}

\newcommand{\cH}{{\mathcal{H}}}

\newcommand{\cO}{{\mathcal{O}}}
\newcommand{\cN}{{\mathcal{N}}}
\newcommand{\cQ}{{\mathcal{Q}}}
\newcommand{\cF}{{\mathcal{F}}}

\newcommand{\indep}{\rotatebox[origin=c]{90}{$\models$}}
\newcommand{\E}{\mathbb{E}}

\newcommand{\anon}{1}

\begin{document}

\def\spacingset#1{\renewcommand{\baselinestretch}%
{#1}\small\normalsize} \spacingset{1}


\if1\anon
{
  \title{\bf Proximal Path-Specific Inference}
  \author{Yang Bai\\
    Department of Statistics and Data Science, National University of Singapore\\
    Sihan Wu \\
    Center for Data Science, Zhejiang University\\
    Baoluo Sun\\
    Department of Statistics and Data Science, National University of Singapore\\
    Yifan Cui\\
    Center for Data Science, Zhejiang University}
    \date{}
  \maketitle
} \fi

\if0\anon
{
  \bigskip
  \bigskip
  \bigskip
  \begin{center}
    {\LARGE\bf Proximal Path-Specific Inference}
\end{center}
  \date{}
  \medskip
} \fi

\bigskip
\begin{abstract}
 Causal mediation analysis has been extended to estimate path-specific effects with multiple intermediate variables, isolating treatment effects through a mediator of interest while excluding pathways through its ancestors. Such analyses address bias from recanting witnesses, i.e., treatment-induced mediator-outcome confounders. However, existing methods typically rely on stringent assumptions precluding general unmeasured confounding, which are often violated in practice. In this paper, we relax these restrictions by leveraging observed covariates as proxy variables to accommodate unmeasured confounding among the treatment, recanting witness, mediator, and outcome. Using proximal confounding bridge functions, we develop four nonparametric identification strategies for the path-specific effect. We further derive the efficient influence function and propose a quadruply robust, locally efficient estimator. To handle high-dimensional nuisance parameters, we propose a proximal debiased machine learning approach. We theoretically guarantee that our estimator achieves $\sqrt{n}$-consistency and asymptotic normality even when machine learning estimators for nuisance functions converge at slower rates. Our approaches are validated via semiparametric and nonparametric simulations and an application to the CDC WONDER Natality study, estimating the path-specific effect of prenatal care on preterm birth through preeclampsia, independent of maternal smoking during pregnancy.
\end{abstract}

\noindent%
\vfill

\newpage
\spacingset{1.8} 

\section{Introduction}\label{sec: introduction}

Mediation analysis \citep{robins1992identifiability, Pearl2001direct, imai2010general, tchetgen2012semiparametric} provides a principled framework for investigating causal mechanisms by decomposing the effect of a treatment $A$ on an outcome $Y$ into pathways operating through a mediator of interest $M$. Classical mediation analysis focuses on the natural indirect effect, corresponding to the pathway from $A$ to $Y$ through $M$, and the natural direct effect, corresponding to pathways not through $M$. These estimands are well understood when a single mediator is present and strong identification assumptions hold.

However, in many applications, there exist multiple intermediate variables between treatment and outcome. In such settings, conventional mediation analysis typically requires the absence of treatment-induced mediator-outcome confounders---often referred to as \textit{recanting witnesses}---as well as the absence of unmeasured confounding. Under these circumstances, commonly used identification assumptions such as sequential ignorability \citep{imai2010identification} or nonparametric structural equation models with independent errors (NPSEM-IE) \citep{pearl2009causality} no longer suffice to identify natural indirect effects \citep{Avin2005identifiability, tchetgen2014identification}. Figure~\ref{fig: causal diagrams} illustrates this issue: the recanting witness $D$ is directly affected by $A$ and simultaneously confounds the relationship between $M$ and $Y$. Such treatment-induced confounding is common in epidemiologic studies, particularly when the mediator of interest occurs long after the treatment initiation \citep{robins1999testing}.

A motivating example arises in studies of preterm birth. Mediation analysis has been widely used to explore whether adequate prenatal care ($A$) reduces the risk of preterm birth ($Y$) through preeclampsia ($M$) \citep{vansteelandt2012natural, vanderweele2014effect, xia2023identification}. In this context, maternal smoking during pregnancy ($D$) acts as a treatment-induced confounder of the relationship between preeclampsia and preterm birth. The presence of such a recanting witness invalidates standard identification of mediation estimands.

One possible solution is to treat smoking and preeclampsia as joint mediators \citep{vanderweele2014effect}. However, the resulting joint indirect effect through $\left \{ D, M \right \}$ conflates fundamentally different social and biological mechanisms, obscuring the specific role of preeclampsia. Moreover, even under additional stringent assumptions regarding no direct or indirect heterogeneity that allow identification of the natural indirect effect through $M$ alone \citep{xia2023identification}, this estimand generally combines multiple causal pathways, including both $A \rightarrow M \rightarrow Y$ and $A \rightarrow D \rightarrow M \rightarrow Y$. As a result, the mediated effect of $M$ may be masked or reversed by pathways involving $D$. For instance, while adequate prenatal care may reduce preterm birth risk by mitigating preeclampsia, it may also prompt smoking cessation, whose short-term withdrawal symptoms could deteriorate preeclampsia and then increase preterm birth risk \citep{vansteelandt2012natural}. If the latter pathway dominates, the estimated natural indirect effect may misleadingly suggest that adequate prenatal care increases preterm birth risk through preeclampsia, although the effect of the causal pathway $A \rightarrow M \rightarrow Y$ is essentially opposite. Such paradoxical findings have been reported in prior analyses of real data \citep{vanderweele2014effect, xia2023identification}, highlighting the limitations of conventional mediation estimands.

When scientific interest centers on the indirect effect operating through a specific mediator $M$, pathways involving a recanting witness $D$ are typically not of interest. In such cases, the natural indirect effect is not an appropriate estimand. Alternative causal mediation estimands have therefore been proposed, most notably path-specific effects and interventional effects \citep{vanderweele2014effect}. The interventional indirect effect \citep{vanderweele2014effect, diaz2021nonparametric} is an analog of the natural indirect effect that substitutes the potential mediator with a random draw, which is independent of the potential outcome, from the distribution of the mediator among the non-treated, while path-specific effects explicitly isolate causal pathways of interest. \cite{Avin2005identifiability} and \cite{shpitser2013counterfactual} established identification conditions for path-specific effects. Semiparametric inference for the path-specific effect of primary interest that operates through a specific mediator while excluding pathways through treatment-induced confounders was subsequently developed by \cite{miles2017quantifying, miles2020semiparametric}. More recently, \cite{liu2024general} proposed nonparametric double machine learning estimators based on the efficient influence function to correct for the first-order bias of plug-in machine learning estimators. However, existing approaches generally rely on stringent assumptions regarding unmeasured confounding. In particular, \cite{miles2017quantifying, miles2020semiparametric} allow unmeasured confounders only in the $D-Y$ relationship, thereby excluding confounding that directly affects the treatment $A$ or the mediator $M$ (Figure~\ref{fig: U in D-Y}). Other recent work \citep{shan2025nonparametric} focuses on missing covariates, indexed by a missingness indicator $R$ ($R=1$ if covariates are fully observed and $R=0$ otherwise), rather than addressing genuinely unmeasured confounders. In the preceding example of preterm birth, the exposures to violence, racism, and discrimination are potential unobserved confounders that are difficult to measure accurately and may directly influence prenatal care utilization $A$, smoking during pregnancy $D$, preeclampsia $M$, and preterm birth $Y$. Consequently, methodological tools for estimating path-specific effects in the presence of recanting witnesses and general unmeasured confounding (Figure~\ref{fig: U among A, D, M, Y}) remain limited.

To address unmeasured confounding in causal inference, the recently developed proximal framework \citep{miao2018identifying, cui2024semiparametric} leverages observed covariates as proxy variables for latent confounders to estimate the average treatment effect. Proximal inference has been extended to a wide range of settings, including longitudinal studies \citep{tchetgen2020introduction, ying2023proximal}, survival analysis \citep{ying2022proximal, ying2024proximal}, individualized and dynamic treatment regimes \citep{shen2023optimal, qi2024proximal, zhang2024identification, gao2025multiple}, continuous treatments \citep{wu2023doubly}, and synthetic control methods for comparative case studies \citep{liu2024proximal, shi2026theory}. Notably, proximal methods have been developed for mediation analysis in the absence of recanting witnesses, enabling identification of natural indirect effect \citep{dukes2023proximal} and population intervention indirect effect \citep{bai2025proximal} under unmeasured confounding, as well as mediated effects with unobserved mediators \citep{ghassami2025causal}. However, proximal methods for path-specific effects in the presence of both recanting witnesses and general unmeasured confounding have not yet been established.

In this paper, we study the causal pathway $A \rightarrow M \rightarrow Y$, which excludes the pathway $A \rightarrow D \rightarrow M \rightarrow Y$, in the presence of treatment-induced mediator-outcome confounder $D$ and unmeasured confounder $U$ that directly affects $A$, $D$, $M$, and $Y$. Our goal is to identify and estimate the path-specific effect through $M$ but not via the recanting witness $D$. Leveraging observed covariates as a pair of proxy variables, we establish sufficient conditions for nonparametric identification of the path-specific effect via proximal confounding bridge functions. We propose four distinct identification strategies based on different combinations of outcome and treatment confounding bridge functions. We further characterize the semiparametric efficiency bound and derive the efficient influence function. Building on these results, we develop a quadruply robust and locally efficient estimator, which remains consistent provided that at least one of the four bridge function combinations is correctly specified, and attains local efficiency when all confounding bridge functions are correctly specified. A notable advantage of using multiply robust estimators based on the efficient influence function is that their second-order bias properties permit slower convergence rates for the estimation of nuisance parameters \citep{chernozhukov2018double, kennedy2024semiparametric, li2025identification}. Furthermore, we propose a proximal debiased machine learning approach based on the derived efficient influence function in conjunction with the cross-fitting technique \citep{schick1986asymptotically, chernozhukov2018double} and provide a theoretical guarantee that $\sqrt{n}$-consistency and asymptotic normality can still be achieved in mild regularity conditions, even when the machine learning methods for estimating nuisance bridge functions may not converge at the parametric $\sqrt{n}$-rate in high-dimensional settings.

We evaluate the proposed methods through two simulation studies—covering both semiparametric and nonparametric scenarios—and demonstrate their practical utility using data from the CDC WONDER Natality study \citep{vansteelandt2012natural}. Specifically, we investigate the mediated effect of prenatal care on preterm birth, identifying preeclampsia as the intermediate pathway. Existing analyses report that the natural indirect effect of adequate prenatal care through preeclampsia increases the risk of preterm birth \citep{vanderweele2014effect, xia2023identification}, a finding that is not entirely unexpected given that improved prenatal care may induce smoking cessation, whose withdrawal symptoms could exacerbate preeclampsia \citep{vansteelandt2012natural}. However, from a clinical and public health perspective, practitioners are primarily interested in whether adequate prenatal care would decrease the risk of preterm birth through preeclampsia itself, excluding pathways operating through smoking during pregnancy. This setting is further complicated by unmeasured confounding, such as exposure to violence, racism, and discrimination, that affects prenatal care utilization, smoking behavior, preeclampsia, and preterm birth but is not recorded in the CDC WONDER Natality dataset. Through both simulation studies and real data analysis, we demonstrate that the proposed methods yield robust and interpretable estimates of path-specific effects under realistic data-generating mechanisms.

\section{Causal estimand, proxy variables, and identification}\label{sec: identification}

In this section, we develop four nonparametric identification strategies for the path-specific effect that operates through the mediator of interest while excluding pathways through treatment-induced confounders, in the presence of recanting witnesses and general unmeasured confounding. Let $A \in \{0,1\}$ denote a binary treatment, $Y$ a continuous or binary outcome, $M$ (possibly multivariate) mediator of interest, and $X$ a vector of observed covariates. We additionally consider a vector of treatment-induced mediator-outcome confounders $D$, commonly referred to as \emph{recanting witnesses} in mediation analysis, due to its involvement in two conflicting causal pathways from $A$ to $Y$, namely $A \rightarrow D \rightarrow M \rightarrow Y$ and $A \rightarrow D \rightarrow Y$, one involving the mediator of interest and the other bypassing it. Furthermore, we let $U$ denote a (possibly multivariate) vector of unmeasured confounders, representing latent social, behavioral, or biological factors, that may simultaneously affect $A$, $D$, $M$, and $Y$. Figure~\ref{fig: U among A, D, M, Y} illustrates this generalized setting, where we focus on the effect of $A$ on $Y$ along the pathway $A \rightarrow M \rightarrow Y$ that excludes the path through $D$. 

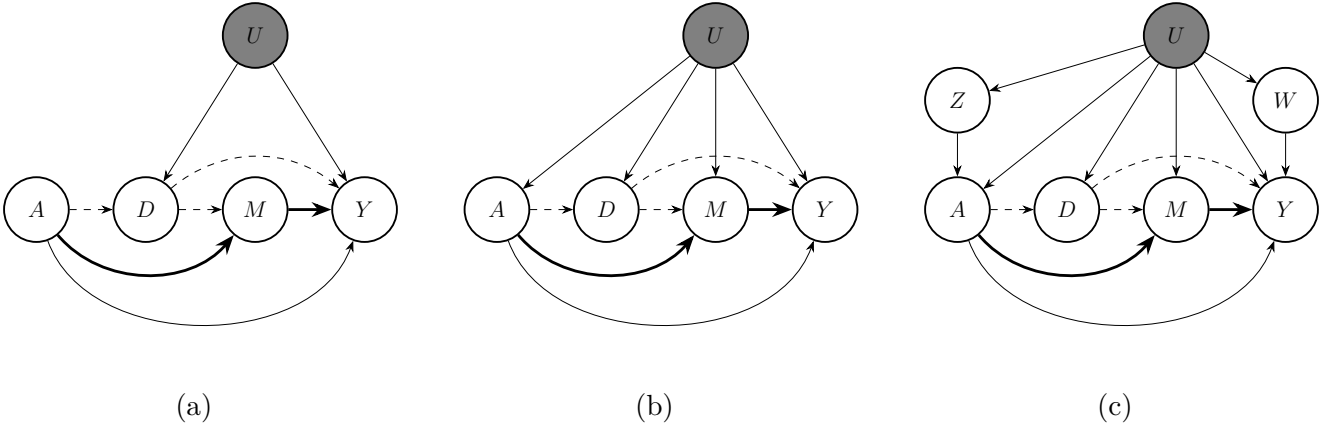
\begin{figure}[ht]
    \centering
    \subfloat[\label{fig: U in D-Y}]{%
        \resizebox{0.3\textwidth}{!}{%
            \begin{tikzpicture}[
                node distance=0.8cm,
                thick,
                normal node/.style={circle, draw, minimum size=1.2cm, font=\sffamily\bfseries},
                hidden node/.style={circle, draw, minimum size=1.2cm, font=\sffamily\bfseries, fill=gray},
                normal edge/.style={->, >=Stealth, thin},
                highlight edge/.style={->, >=Stealth, very thick}
            ]
                \node[normal node] (A) {$A$};
                \node[normal node, right=of A] (D) {$D$};
                \node[normal node, right=of D] (M) {$M$};
                \node[normal node, right=of M] (Y) {$Y$};
                \node[hidden node, above=2cm of M] (U) {$U$};
                
                \draw[normal edge, dashed] (A) to (D);
                \draw[normal edge, dashed] (D) to (M);
                \draw[highlight edge] (M) to (Y);
                \draw[normal edge] (U) to (D);
                \draw[normal edge] (U) to (Y);
                \draw[normal edge, dashed] (D) to[out=40,in=140] (Y);
                \draw[highlight edge] (A) to[out=310,in=230] (M);
                \draw[normal edge] (A) to[out=290,in=250] (Y);
            \end{tikzpicture}%
        }%
    }\hfill%
    \subfloat[\label{fig: U among A, D, M, Y}]{%
        \resizebox{0.3\textwidth}{!}{%
            \begin{tikzpicture}[
                node distance=0.8cm,
                thick,
                normal node/.style={circle, draw, minimum size=1.2cm, font=\sffamily\bfseries},
                hidden node/.style={circle, draw, minimum size=1.2cm, font=\sffamily\bfseries, fill=gray},
                normal edge/.style={->, >=Stealth, thin},
                highlight edge/.style={->, >=Stealth, very thick}
            ]
                \node[normal node] (A) {$A$};
                \node[normal node, right=of A] (D) {$D$};
                \node[normal node, right=of D] (M) {$M$};
                \node[normal node, right=of M] (Y) {$Y$};
                \node[hidden node, above=2cm of M] (U) {$U$};

                \draw[normal edge, dashed] (A) to (D);
                \draw[normal edge, dashed] (D) to (M);
                \draw[highlight edge] (M) to (Y);
                \draw[normal edge] (U) to (A);
                \draw[normal edge] (U) to (D);
                \draw[normal edge] (U) to (M);
                \draw[normal edge] (U) to (Y);
                \draw[normal edge, dashed] (D) to[out=40,in=140] (Y);
                \draw[highlight edge] (A) to[out=310,in=230] (M);
                \draw[normal edge] (A) to[out=290,in=250] (Y);
            \end{tikzpicture}%
        }%
    }\hfill%
    \subfloat[\label{fig: proximal framework}]{%
        \resizebox{0.3\textwidth}{!}{%
            \begin{tikzpicture}[
                node distance=0.8cm,
                thick,
                normal node/.style={circle, draw, minimum size=1.2cm, font=\sffamily\bfseries},
                hidden node/.style={circle, draw, minimum size=1.2cm, font=\sffamily\bfseries, fill=gray},
                normal edge/.style={->, >=Stealth, thin},
                highlight edge/.style={->, >=Stealth, very thick}
            ]
                \node[normal node] (A) {$A$};
                \node[normal node, right=of A] (D) {$D$};
                \node[normal node, right=of D] (M) {$M$};
                \node[normal node, right=of M] (Y) {$Y$};
                \node[hidden node, above=2cm of M] (U) {$U$};
                \node[normal node, above=of A] (Z) {$Z$};
                \node[normal node, above=of Y] (W) {$W$};
                
                \draw[normal edge, dashed] (A) to (D);
                \draw[normal edge, dashed] (D) to (M);
                \draw[highlight edge] (M) to (Y);
                \draw[normal edge] (U) to (A);
                \draw[normal edge] (U) to (D);
                \draw[normal edge] (U) to (M);
                \draw[normal edge] (U) to (Y);
                \draw[normal edge] (U) to (W);
                \draw[normal edge] (W) to (Y);
                \draw[normal edge] (U) to (Z);
                \draw[normal edge] (Z) to (A);
                \draw[normal edge, dashed] (D) to[out=40,in=140] (Y);
                \draw[highlight edge] (A) to[out=310,in=230] (M);
                \draw[normal edge] (A) to[out=290,in=250] (Y);
            \end{tikzpicture}%
        }%
    }
    \caption{Comparison of causal diagrams for path-specific effects: (a) unmeasured confounding restricted to $D-Y$ considered by \cite{miles2017quantifying, miles2020semiparametric}; (b) generalized unmeasured confounding across all variables; and (c) the proposed proximal framework with proxies $Z$ and $W$. Gray nodes denote unmeasured variables $U$, dashed arrows denote the role of \textit{recanting witnesses} $D$, and bold paths indicate the effect $\mathcal{P}_{AMY}$. Observed covariates $X$ are omitted for clarity.}
    \label{fig: causal diagrams}
\end{figure}

To formalize the path-specific effect of interest, we introduce the counterfactual framework \citep{rubin1974estimating}. Let $Y(a, d, m)$ denote the potential outcome that would be observed if, possibly contrary to fact, the treatment $A$, recanting witness $D$, and mediator $M$ were assigned to a given level $a$, $d$, and $m$, respectively. In mediation analysis, there are also counterfactuals for intermediate variables, and we define $D(a)$, $M(a)$, $M(a,d)$, $Y(a)$, and $Y(a,m)$ analogously. The path-specific effect of $A$ on $Y$ along the path $A \rightarrow M \rightarrow Y$ with respect to comparing the treatment level $A=1$ with the control level $A=0$ on the mean difference scale is defined in terms of the difference in expectations of two nested counterfactual outcomes,
\begin{equation}
    \mathcal{P}_{AMY} = \E\left [ Y\left ( 1, D(1), M\left ( 1, D(1) \right ) \right ) \right ] - \E\left [ Y\left ( 1, D(1), M\left ( 0, D(1) \right ) \right ) \right ].
\end{equation}
This contrast isolates the causal effect of $A$ on $Y$ transmitted through $M$ while holding the recanting witness $D$ fixed at its natural level under $A=1$, thereby excluding all pathways that operate through $D$. Note that the first term $\E\left [ Y\left ( 1, D(1), M\left ( 1, D(1) \right ) \right ) \right ] = \E\left [ Y\left ( 1 \right ) \right ]$ is the counterfactual mean outcome in the absence of intermediate variables; this term is identified under the unmeasured confounding condition, $Y(a) \indep A \mid U, X$, in the proximal inference framework \citep{miao2018identifying, cui2024semiparametric} on the average treatment effect, which holds in Figure~\ref{fig: causal diagrams}. We therefore treat the first term as known up to estimation error and focus on the second term, denoted by $\psi = \E\left [ Y\left ( 1, D(1), M\left ( 0, D(1) \right ) \right ) \right ]$. The mediation functional $\psi$ represents the mean potential outcome under an intervention that sets $A=1$ and allows the recanting witness to follow its natural value $D(1)$, while forcing the mediator to take the value it would have attained under $A=0$ given $D(1)$. In contrast, the natural indirect effect, $NIE = \E[Y(1,M(1))] - \E[Y(1,M(0))]$, conflates the pathways $A \rightarrow M \rightarrow Y$ and $A \rightarrow D \rightarrow M \rightarrow Y$.

To account for the unmeasured confounding in identifying $\psi$, \cite{miles2017quantifying, miles2020semiparametric} consider an ideal circumstance (Figure~\ref{fig: U in D-Y}) where one can access abundant observed covariates $X$ such that the remaining unmeasured covariates $U$ only confounds the $D-Y$ relationship instead of directly affecting $A$ or $Y$, and impose the conditional exchangeability assumptions: (i) $\left \{ Y(a,m), D(a) \right \} \indep A \mid X$ for $a \in \left \{ 0,1 \right \}$ and all $m$; (ii) $Y(m) \indep M \mid D, A, X$ for all $m$; (iii) $M(a,d) \indep \left \{ A,D \right \} \mid X$ for $a \in \left \{ 0,1 \right \}$ and all $d$; and (iv) $\left \{ Y(a,m), D(a) \right \} \indep M(a^{\ast},d) \mid X$ for $a, a^{\ast} \in \left \{ 0,1 \right \}$, all $d$ and all $m$. The first conditional independence requires that the effects of $A$ on $D$ and $Y$ are unconfounded conditional on $X$; the second states that the effect of $M$ on $Y$ is unconfounded conditional on $D$, $A$, and $X$; the third imposes that the effects of $A$ and $D$ on $M$ are unconfounded conditional on $X$; and the fourth is a cross-world counterfactual condition that states $\left \{ Y(a,m), D(a) \right \}$ under the intervention $A=a$ is independent of $M(a^{\ast},d)$ from another different world with the intervention $A=a^{\ast}$ conditional on $X$, which cannot be empirically verified and is typically justified under an NPSEM-IE interpretation \citep{robins2010alternative}. If (i)-(iv) hold, then $\psi$ can be identified by the following formula \citep{miles2017quantifying, miles2020semiparametric}:
\begin{equation}
    \psi = \int\int\int \E\left [ Y \mid A=1, d, m, x \right ]\mathrm{d}F_{M}(m|A=0, d, x) \mathrm{d}F_{D}(d|A=1, x) \mathrm{d}F_{X}(x).
  \label{eq_id unconfounded}
\end{equation}
If Figure~\ref{fig: U in D-Y} is interpreted in the sense of NPSEM-IE, then the conditional exchangeability assumptions~(i)-(iv) are consistent with this diagram, and the identification formula~\eqref{eq_id unconfounded} holds.

However, in most applications, it is unrealistic to observe a sufficiently rich set of covariates to account for the probably infinite unmeasured confounders. In the earlier example of preterm birth, the unmeasured confounder $U$ may be exposures to violence and racism that are not recorded in epidemiological studies \citep{vanderweele2014effect, xia2023identification} and directly affect prenatal care utilization $A$, maternal smoking during pregnancy $D$, preeclampsia $M$, and preterm birth $Y$. Figure~\ref{fig: U among A, D, M, Y} illustrates a more realistic scenario, where the conditional exchangeability assumptions~(i)-(iv) are violated and identification~\eqref{eq_id unconfounded} fails due to the presence of a general unmeasured confounder $U$ that is a common cause of $A$, $D$, $M$, and $Y$. Before giving our identification results, we formalize this general circumstance by introducing the following assumptions.

\begin{assumption}(Consistency):
    (i) $D(a) = D$ almost surely if $A=a$; 
    (ii) $M(a,d) = M$ almost surely if $A=a$ and $D=d$; 
    (iii) $Y(a,d,m) = Y$ almost surely if $A=a$, $D=d$, and $M=m$.
  \label{asm: consistency}
\end{assumption}

\begin{assumption}(Positivity): 
    For $a=0, 1$,  and all $\left \{ d, m \right \}$, we have $\mathbb{P}(A=a \mid U, X) >0$, $f_{D|A,U,X}(d \mid A, U, X) >0$, and $f_{M|D,A,U,X}(m \mid D, A, U, X) >0$ almost surely.
  \label{asm: positivity}
\end{assumption}

\begin{assumption}(Latent conditional exchangeability): 
    For $a=0, 1$,  and all $\left \{ d, m \right \}$, we have (i) $\left \{ Y(a,m), D(a) \right \} \indep A \mid U, X$; 
    (ii) $Y(m) \indep M \mid D, A, U, X$; 
    and (iii) $M(a,d) \indep \left \{ D, A \right \} \mid U, X$.
  \label{asm: latent exchangeability}
\end{assumption}

\begin{assumption}(Latent cross-world independence): 
    For $a, a^{\ast} = 0, 1$, and all $\left \{ d, m \right \}$, we have $\left \{ Y(a,m), D(a) \right \} \indep M(a^{\ast}, d) \mid U, X$.
  \label{asm: cross-world}
\end{assumption}

Assumptions~\ref{asm: consistency}-\ref{asm: positivity} are standard consistency and positivity conditions imposed to ensure the identification in causal inference \citep{robins1986new}. The consistency assumption~\ref{asm: consistency} is typically needed to make any progress towards linking counterfactuals with the observed variables. The positivity assumption~\ref{asm: positivity} prevents extreme probability weights of treatment, recanting witness, and mediator when leveraging inverse probability weighted estimators. Assumptions~\ref{asm: latent exchangeability}-\ref{asm: cross-world} generalize the standard conditional exchangeability assumptions~(i)-(iv) in the model of \cite{miles2017quantifying, miles2020semiparametric}, by accommodating a general unmeasured confounder $U$ that exerts direct effects on $A$, $D$, $M$, and $Y$. Specifically, the conditional exchangeability assumptions~(i)-(iv) do require that observed covariates $X$ be sufficient to control for confounding of the effects of $A$ on $Y$, $A$ on $D$, $M$ on $Y$, $A$ on $M$, and $D$ on $M$, while Assumption~\ref{asm: latent exchangeability} relaxes this restriction to allow these effects subject to the unmeasured confounding. Assumption~\ref{asm: cross-world} states that the independence between counterfactual outcome, recanting witness, and mediator values across two different worlds is also subject to the unmeasured confounder $U$. Figure~\ref{fig: U among A, D, M, Y} demonstrates Assumptions~\ref{asm: latent exchangeability}-\ref{asm: cross-world} under the NPSEM-IE interpretation of the causal diagram.

The failure of conditional exchangeability~(i)-(iv) indicates that an analysis which is only conditional on the observed covariates would return biased results due to residual unmeasured confounding from $U$. In what follows, we will develop a different approach for the identification of $\psi$. We suppose that some components of observed covariates suffice to learn about the unmeasured confounding, where $Z$ and $W$ are adequate to serve as proxy variables for $U$. Specifically, the treatment-inducing confounding proxy $Z$ is a potential cause of treatment $A$ and is associated with $D$, $M$, and $Y$ only through its effect on $A$ and shared causes $X$ and $U$; the outcome-inducing confounding proxy $W$ is a potential cause of outcome $Y$ and is associated with $A$, $D$, and $M$ only through its shared causes $X$ and $U$. Additionally, $Z$ is associated with $W$ only via their shared causes $X$ and $U$. The following assumptions formalize the properties of proxy variables $Z$ and $W$, and enable us to leverage information from them to address the unmeasured confounding. 

\begin{assumption}(Proxy variables):
    (i) $W \indep \left \{ A, D, M \right \} \mid U, X$;
    (ii) $Z \indep \left \{ W, D, M, Y \right \} \mid A, U, X$.
  \label{asm: proxy variables}
\end{assumption}

Figure~\ref{fig: proximal framework} provides a graphical illustration of Assumption~\ref{asm: proxy variables} that proxy variables $Z$ and $W$ satisfy in the context of the generalized mediation model formalized by Assumptions~\ref{asm: consistency}-\ref{asm: cross-world}. Assumption~\ref{asm: proxy variables} essentially requires that there is no direct effect between $W$ and $\left \{ A, D, M \right \}$, between $Z$ and $\left \{ W, D, M, Y \right \}$, and between $Z$ and $W$. Notably, Figure~\ref{fig: proximal framework} demonstrates only one possible graphical model to be compatible with Assumption~\ref{asm: proxy variables} for proxy variables. There are alternative diagrams to satisfy the proxy variable conditions; for instance, the edges from $Z$ to $A$ and from $W$ to $Y$ could also be reversed, or there could exist extra confounding in the relationship between $Z$ and $A$ or between $W$ and $Y$. Hence, practitioners possess the flexibility to select proper observed covariates as proxy variables for unmeasured confounding in clinical studies.

\begin{assumption}(Completeness$-Z$):
    (i) For any square-integrable function $g(u)$ and any $\left \{ m,d,x \right \}$, if $\E\left [ g(U) \mid Z, A=1, M=m, D=d, X=x \right ] = 0$ almost surely, then $g(U)=0$ almost surely.
    (ii) For any square-integrable function $g(u)$ and any $\left \{ d,x \right \}$, if $\E\left [ g(U) \mid Z, A=0, D=d, X=x \right ] = 0$ almost surely, then $g(U)=0$ almost surely.
    (iii) For any square-integrable function $g(u)$ and any $x$, if $\E\left [ g(U) \mid Z, A=1, X=x \right ] = 0$ almost surely, then $g(U)=0$ almost surely.
  \label{asm: completeness for Z}
\end{assumption}

Completeness is a technical condition linked to sufficiency in statistical inference and is critical for identification in nonparametric regression with instrumental variables \citep{newey2003instrumental}. In our context, Assumption~\ref{asm: completeness for Z} connects the range of $U$ to that of $Z$, ensuring that any variation in $U$ is fully captured by the corresponding variation in $Z$ in three scenarios, including conditional on $\left \{ A=1, M, D, X \right \}$, $\left \{ A=0, D, X \right \}$, or $\left \{ A=1, X \right \}$. For the case of categorical $U$ and $Z$ with number of categories $\textbf{\#}_u$ and $\textbf{\#}_z$, respectively, Assumption~\ref{asm: completeness for Z} essentially implies $\textbf{\#}_z \geq \textbf{\#}_u$,
indicating that proxy variables possess at least as many categories as the unmeasured confounders. For continuously distributed unmeasured confounders $U$, completeness holds under many common distributions. For instance, suppose that the distribution of $U$ is absolutely continuous conditional on $Z$, $A=1$, $D$, $M$, and $X$, with density
\begin{equation}
    f(u|z, A=1, d, m, x) = s(u) \exp\left\{\eta(z, d, m, x)^T t(u) - \alpha(z, d, m, x) \right\},
    \label{eq: exponential family}
\end{equation}
where $s(u)>0$, $t(u)$ is a one-to-one function of $u$, and the support of $\eta(z, d, m, x)$ is an open set; then the completeness assumption~\ref{asm: completeness for Z} holds \citep{newey2003instrumental}. Notably, model~\eqref{eq: exponential family} is the exponential family with parameters $z$, $d$, $m$, and $x$. The justifications for completeness have also been discussed for location-scale families \citep{hu2018nonparametric}, as well as nonparametric instrumental regression models \citep{d2011completeness, darolles2011nonparametric}. Thus, accumulating observed covariates to form a rich collection of proxy candidates offers a better opportunity to control for the unmeasured confounding.

With proper proxy variables $Z$ and $W$, we now propose the first identification result.

\begin{theorem}
    Suppose that there exist the outcome confounding bridge functions $h_2(W, M, D, X)$, $h_1(W, D, X)$, and $h_0(W, X)$, solving the following integral equations:
    \begin{align}
        \E\left ( Y \mid A=1, M, D, Z, X \right ) =& \E\left [ h_2(W, M, D, X) \mid A=1, M, D, Z, X \right ], \label{eq_solving h2} \\
        \E\left [ h_2(W, M, D, X) \mid A=0, D, Z, X \right ] =& \E\left [ h_1(W, D, X) \mid A=0, D, Z, X \right ], \label{eq_solving h1} \\
        \E\left [ h_1(W, D, X) \mid A=1, Z, X \right ] =& \E\left [ h_0(W, X) \mid A=1, Z, X \right ]. \label{eq_solving h0}
    \end{align}
    Under Assumptions~\ref{asm: consistency}-\ref{asm: completeness for Z}, $\psi = \E\left [ Y\left ( 1, D(1), M\left ( 0, D(1) \right ) \right ) \right ]$ is identified by
    \begin{equation}
        \psi = \E\left [ h_0(W, X) \right ]. \label{eq_id h0}
    \end{equation}
 \label{thm-h2h1h0}
\end{theorem}

\vspace{-0.5cm}
The proof of Theorem~\ref{thm-h2h1h0}, as well as other theoretical results, is provided in the supplementary material. Similar to the g-formula developed by \cite{tchetgen2020introduction}, the identification formula~\eqref{eq_id h0} relies on nested confounding bridge functions, which are formalized in \eqref{eq_solving h2}-\eqref{eq_solving h0}. The existence of these inverse problems~\eqref{eq_solving h2}-\eqref{eq_solving h0}, also referred to as the Fredholm integral equation of the first kind, is well elaborated in literature \citep{miao2018identifying, cui2024semiparametric}. Notably, the uniqueness of solutions to \eqref{eq_solving h2}-\eqref{eq_solving h0} is not necessary in the process of identification, because all the solutions induce the unique value of $\psi$ in \eqref{eq_id h0}. Assumptions~\ref{asm: proxy variables}-\ref{asm: completeness for Z} ensure that the proxy variables $Z$ and $W$ sufficiently adjust for unmeasured confounding. Specifically, they guarantee that the outcome confounding bridge functions $h_2$, $h_1$, and $h_0$, though identified via the observed data distribution in \eqref{eq_solving h2}-\eqref{eq_solving h0}, remain valid representations of the underlying latent structures involving $U$. Interestingly, although inference on the path-specific effect $\mathcal{P}_{AMY}$ requires understanding the interventions on recanting witness $D$ and mediator $M$, it can be identified only with the help of treatment-inducing and outcome-inducing proxies $Z$ and $W$ that have no direct effect on $D$ and $M$, without the need for additional proxy variables induced by $D$ or $M$.

As an alternative to Assumption~\ref{asm: completeness for Z} that ensures sufficient variability of $Z$, the following assumption instead enables the variability of $W$ to account for the variability of $U$ sufficiently.

\begin{assumption}(Completeness$-W$):
    (i) For any square-integrable function $g(u)$ and any $x$, if $\E\left [ g(U) \mid W, A=1, X=x \right ] = 0$ almost surely, then $g(U)=0$ almost surely.
    (ii) For any square-integrable function $g(u)$ and any $\left \{ d,x \right \}$, if $\E\left [ g(U) \mid W, A=0, D=d, X=x \right ] = 0$ almost surely, then $g(U)=0$ almost surely.
  \label{asm: completeness for W}
\end{assumption}

Assumption~\ref{asm: completeness for W} states that the variation in $W$ fully captures the variation in $U$ conditional on $\left \{ A=1, X \right \}$ or on $\left \{ A=0, D, X \right \}$. This further implies that $W$ exhibits sufficient variability relative to $U$, conditional on $A$, $D$, $M$, and $X$. With Assumption~\ref{asm: completeness for W} replacing Assumption~\ref{asm: completeness for Z}, we establish the second identification strategy, which relies on different confounding bridge functions as counterparts to $h_2$, $h_1$, and $h_0$.

\begin{theorem}
    Suppose that there exist the treatment confounding bridge functions $q_0(Z, X)$, $q_1(Z, D, X)$, and $q_2(Z, M, D, X)$ that sequentially solve the following integral equations:
    {\small \begin{align}
        \frac{1}{f(A=1 | W, X)} =& \E\left [ q_0(Z, X) \mid A=1, W, X \right ], \label{eq_solving q0} \\
        \E\left [ q_0(Z, X) \mid A=1, D, W, X \right ] \frac{f(A=1|D, W, X)}{f(A=0|D, W, X)} =& \E\left [ q_1(Z, D, X) \mid A=0, D, W, X \right ], \label{eq_solving q1} \\
        \E\left [ q_1(Z, D, X) \mid A=0, M, D, W, X \right ] \frac{f(A=0|M, D, W, X)}{f(A=1|M, D, W, X)} =& \E\left [ q_2(Z, M, D, X) \mid A=1, M, D, W, X \right ]. \label{eq_solving q2}
    \end{align}}
    Under Assumptions~\ref{asm: consistency}-\ref{asm: proxy variables} and \ref{asm: completeness for W}, $\psi = \E\left [ Y\left ( 1, D(1), M\left ( 0, D(1) \right ) \right ) \right ]$ is identified by
    \begin{equation}
        \psi = \E\left [ I(A=1) Y q_2(Z, M, D, X) \right ]. \label{eq_id q2}
    \end{equation}
 \label{thm-q0q1q2}
\end{theorem}

\vspace{-0.5cm}
The identification formula~\eqref{eq_id q2} relies on nested treatment confounding bridge functions $q_0$, $q_1$, and $q_2$ defined by \eqref{eq_solving q0}-\eqref{eq_solving q2} involving propensity scores for treatment, to which the existence of solutions is discussed by \cite{cui2024semiparametric}. In the following result, we note that their solving equations~\eqref{eq_solving q0}-\eqref{eq_solving q2} are equivalent to some conditional moment equations as counterparts to \eqref{eq_solving h2}-\eqref{eq_solving h0}, which we will employ to construct confounding bridge function estimators that essentially avoid estimating those propensity scores in the following sections.

\begin{proposition}
    The treatment confounding bridge functions $q_0$, $q_1$, and $q_2$ sequentially solve integral equations~\eqref{eq_solving q0}-\eqref{eq_solving q2} if and only if they sequentially solve the following integral equations:
    {\small \begin{align}
        \E\left [ A q_0(Z,X) -1 \mid W, X \right ] =& 0 \label{eq_solving q0 alter} \\
        \E\left [ (1-A) q_1(Z, D, X) - A q_0(Z,X) \mid W, D, X \right ] =& 0 \label{eq_solving q1 alter} \\
        \E\left [ A q_2(Z, M, D, X) - (1-A) q_1(Z, D, X) \mid W, M, D, X \right ] =& 0. \label{eq_solving q2 alter}
    \end{align}}
  \label{propo-solving q0 q1 q2}
\end{proposition}

\vspace{-0.5cm}
Theorem~\ref{thm-h2h1h0} achieves the identification of $\psi$ relying on three nested outcome confounding bridge functions, which are obtained in the order of $h_2$, $h_1$, and $h_0$ by solving integral equations~\eqref{eq_solving h2}-\eqref{eq_solving h0}. The following theorem gives a strategy to overcome the dilemma where the solution of $h_0$ to \eqref{eq_solving h0} is unavailable while $h_2$ and $h_1$ are gained, which will rely on the availability of an additional treatment confounding bridge function $q_0$ instead.

\begin{theorem}
    Suppose that there exist two outcome confounding bridge functions $h_2(W, M, D, X)$ and $h_1(W, D, X)$ that satisfy the integral equations~\eqref{eq_solving h2} and \eqref{eq_solving h1}, as well as one treatment confounding bridge function $q_0(Z, X)$ that solves the integral equation~\eqref{eq_solving q0}. Then, under Assumptions~\ref{asm: consistency}-\ref{asm: proxy variables}, \ref{asm: completeness for Z}\textcolor{blue}{(i)}, \ref{asm: completeness for Z}\textcolor{blue}{(ii)}, and \ref{asm: completeness for W}\textcolor{blue}{(i)}, $\psi = \E\left [ Y\left ( 1, D(1), M\left ( 0, D(1) \right ) \right ) \right ]$ is identified by
    \begin{equation}
        \psi = \E\left [ I(A=1) h_1(W, D, X) q_0(Z, X) \right ].  \label{eq_id h1q0}
    \end{equation}
 \label{thm-h2h1q0}
\end{theorem}

\vspace{-0.5cm}
Analogously, Theorem~\ref{thm-q0q1q2} achieves the identification of $\psi$ relying on three nested treatment confounding bridge functions, which are obtained in the order of $q_0$, $q_1$, and $q_2$ by solving integral equations~\eqref{eq_solving q0}-\eqref{eq_solving q2}. The following theorem gives a strategy to overcome the dilemma where the solution of $q_2$ to \eqref{eq_solving q2} is unavailable while $q_0$ and $q_1$ are gained, which will rely on the availability of an additional outcome confounding bridge function $h_2$ instead.

\begin{theorem}
    Suppose that there exist one outcome confounding bridge function $h_2(W, M, D, X)$ satisfying the integral equation~\eqref{eq_solving h2}, and also two treatment confounding bridge functions $q_0(Z, X)$ and $q_1(Z, D, X)$ solving the integral equations~\eqref{eq_solving q0} and \eqref{eq_solving q1}. Then, under Assumptions~\ref{asm: consistency}-\ref{asm: proxy variables}, \ref{asm: completeness for Z}\textcolor{blue}{(i)}, and \ref{asm: completeness for W}, $\psi = \E\left [ Y\left ( 1, D(1), M\left ( 0, D(1) \right ) \right ) \right ]$ is identified by
    \begin{equation}
        \psi = \E\left [ I(A=0) h_2(W, M, D, X) q_1(Z, D, X) \right ]. \label{eq_id h2q1}
    \end{equation}
 \label{thm-h2q0q1}
\end{theorem}

\vspace{-0.5cm}
In summary, we establish four complementary strategies for the identification of $\psi$, each of which relies on three confounding bridge functions. Specifically, Theorem~\ref{thm-h2h1h0} relies on three outcome confounding bridge functions $\left \{ h_2, h_1, h_0 \right \}$, Theorem~\ref{thm-q0q1q2} relies on three treatment confounding bridge functions $\left \{ q_0, q_1, q_2 \right \}$, while Theorems~\ref{thm-h2h1q0} and \ref{thm-h2q0q1} rely on a hybrid combination of outcome and treatment confounding bridge functions $\left \{ h_2, h_1, q_0 \right \}$ and $\left \{ q_0, q_1, h_2 \right \}$, respectively.

\section{Semiparametric quadruply robust estimation}\label{sec: semipara}

In this section, we develop semiparametric estimators for the parameter of interest based on its efficient influence function. We show how the influence function representation yields a quadruply robust estimator that remains consistent under a union of four semiparametric models. We further establish local efficiency at their intersection.

While the complementary identification strategies~\eqref{eq_id h0}, \eqref{eq_id q2}, \eqref{eq_id h1q0}, and \eqref{eq_id h2q1} proposed in Theorems~\ref{thm-h2h1h0}, \ref{thm-q0q1q2}, \ref{thm-h2h1q0}, and \ref{thm-h2q0q1} can be directly used to construct estimators of $\psi$, such estimators generally rely on the correct specification of nuisance functions and may incur first-order bias when these nuisance functions are estimated using flexible or machine-learning methods. To address these limitations, we instead base estimation on the efficient influence function of $\psi$ under a semiparametric model $\cM_{sp}$, which places no restrictions on the observed data distribution beyond the existence (but not uniqueness) of outcome confounding bridge functions $h_2$, $h_1$, and $h_0$ solving \eqref{eq_solving h2}-\eqref{eq_solving h0}. The resulting estimator is quadruply robust against model misspecification, locally efficient when all nuisance functions are consistently estimated, and exhibits second-order bias with respect to nuisance estimation error \citep{robins2008higher, robins2017minimax}.

Let $T_2: L_2(W, M, D, X) \rightarrow L_2(Z, A=1, M, D, X)$ represent the conditional expectation operator such that $T_2(g) = \E\left [ g(W, M, D, X) \mid Z, A=1, M, D, X \right ]$; $T_1: L_2(W, M, D, X) \rightarrow L_2(Z, A=0, D, X)$ such that $T_1(g) = \E\left [ g(W, M, D, X) \mid Z, A=0, D, X \right ]$; and $T_0: L_2(W, D, X) \rightarrow L_2(Z, A=1, X)$ such that $T_0(g) = \E\left [ g(W, D, X) \mid Z, A=1, X \right ]$. The existence of confounding bridge functions imposes restrictions on the tangent space of $\cM_{sp}$, which allows us to characterize the semiparametric efficiency bound under an additional regularity condition.

\begin{assumption}(Regularity): $T_2$, $T_1$, and $T_0$ are surjective at the true data-generating mechanism.
    \label{asm: regularity}
\end{assumption}

\vspace{-0.3cm}
Assumption~\ref{asm: regularity} states that functions in $L_2(W, M, D, X)$ and $L_2(W, D, X)$ are sufficiently abundant such that spaces $L_2(Z, A=1, M, D, X)$, $L_2(Z, A=0, D, X)$, and $L_2(Z, A=1, X)$ can be generated through conditional expectation operations.

\begin{theorem}
    Assume the existence of outcome confounding bridge functions $h_2$, $h_1$, and $h_0$ in $\cM_{sp}$, and treatment confounding bridge functions $q_0$, $q_1$, and $q_2$ solving \eqref{eq_solving q0}-\eqref{eq_solving q2} at the true data-generating law. If Assumptions~\ref{asm: consistency}-\ref{asm: completeness for Z} hold so that $\psi$ is uniquely identified, then
    {\small \begin{align*}
        EIF_{\psi} =& I(A=1) q_0(Z,X) \left [ h_1(W,D,X) - h_0(W,X) \right ]
        + I(A=0) q_1(Z,D,X) \left [ h_2(W,M,D,X) - h_1(W,D,X) \right ] \\
        &+ I(A=1) q_2(Z,M,D,X) \left [ Y - h_2(W,M,D,X) \right ] + h_0(W,X) - \psi
    \end{align*}}
    is a valid influence function for the estimation of $\psi$ under $\cM_{sp}$. Furthermore, it is also the efficient influence function for $\psi$ at the submodel where confounding bridge functions $\left \{ h_2, h_1, h_0 \right \}$ and $\left \{ q_0, q_1, q_2 \right \}$ are uniquely identified and Assumption~\ref{asm: regularity} holds. The corresponding semiparametric efficiency bound of $\psi$ is given by $\E(EIF_{\psi}^2)$.
 \label{thm-EIF}
\end{theorem}

Notably, Assumption~\ref{asm: regularity} and uniqueness of confounding bridge functions are only required to hold for the local efficiency statement. In other words, the asymptotic variance of any regular and asymptotically linear (RAL) estimator \citep{tsiatis2006semiparametric} of $\psi$ is no smaller than the efficiency bound $\E(EIF_{\psi}^2)$ described in Theorem~\ref{thm-EIF}. An immediate consequence of the influence function representation is an alternative identification formula for $\psi$, which does not rely on the regularity condition or uniqueness of confounding bridge functions.

\begin{corollary}
    Suppose that there exist outcome confounding bridge functions $h_2(W, M, D, X)$, $h_1(W, D, X)$, and $h_0(W, X)$ that sequentially solve \eqref{eq_solving h2}-\eqref{eq_solving h0}, and treatment confounding bridge functions $q_0(Z, X)$, $q_1(Z, D, X)$, and $q_2(Z, M, D, X)$ that sequentially solve \eqref{eq_solving q0}-\eqref{eq_solving q2}. Then, under Assumptions~\ref{asm: consistency}-\ref{asm: completeness for W}, $\psi$ can be nonparametrically identified by
    \begin{align}
        \psi = \E&\left\{ A q_0(Z,X) \left [ h_1(W,D,X) - h_0(W,X) \right ]
        + (1-A) q_1(Z,D,X) \left [ h_2(W,M,D,X) - h_1(W,D,X) \right ] \right. \notag \\
        &+ \left. (1-A) q_2(Z,M,D,X) \left [ Y - h_2(W,M,D,X) \right ] + h_0(W,X) \right\}.
      \label{eq_id robust}
    \end{align}
 \label{coro-id robust}
\end{corollary}

\vspace{-0.7cm}
The identification strategy~\eqref{eq_id robust} established in Corollary~\ref{coro-id robust} can also be leveraged to design estimators of $\psi$. In the following, we will elaborate on the quadruple robustness of \eqref{eq_id robust} against model misspecification, which is one important superiority over \eqref{eq_id h0}, \eqref{eq_id q2}, \eqref{eq_id h1q0}, and \eqref{eq_id h2q1} to construct a consistent estimator. The proposed approaches rely on models for estimating confounding bridge functions. To make the elaboration more concrete, we consider four semiparametric models, each corresponding to a distinct identification strategy and requiring the correct specification of different subsets of confounding bridge functions:
\begin{align*}
    \cM_1:& h_2(W,M,D,X), h_1(W,D,X), \text{and}~ h_0(W,X) ~\text{are correctly specified}. \\
    \cM_2:& q_0(Z,X), q_1(Z,D,X), \text{and}~ q_2(Z,M,D,X) ~\text{are correctly specified}.  \\
    \cM_3:& h_2(W,M,D,X), h_1(W,D,X), \text{and}~ q_0(Z,X) ~\text{are correctly specified}. \\
    \cM_4:& h_2(W,M,D,X), q_0(Z,X), \text{and}~ q_1(Z,D,X) ~\text{are correctly specified}.
\end{align*}
Here, $\cM_1$, $\cM_2$, $\cM_3$, and $\cM_4$ characterize the corresponding identification conditions that guarantee \eqref{eq_id h0}, \eqref{eq_id q2}, \eqref{eq_id h1q0}, and \eqref{eq_id h2q1} to hold, respectively. Correctly specifying different confounding functions to satisfy a specific model among $\left \{ \cM_1, \cM_2, \cM_3, \cM_4 \right \}$ may be challenging, since they are defined as solutions to integral equations. Hence, the development of quadruply robust estimation is helpful, ensuring that when any one of $\left \{ \cM_1, \cM_2, \cM_3, \cM_4 \right \}$ is satisfied but unnecessarily known which one is satisfied, the identification strategy~\eqref{eq_id robust} still holds.

Before expounding on the quadruple robustness, we first consider the way to obtain semiparametric inference for nuisance functions in submodels $\cM_1$, $\cM_2$, $\cM_3$, and $\cM_4$. Let $h_2(W,M,D,X;\beta_2)$, $h_1(W,D,X;\beta_1)$, and $h_0(W,X;\beta_0)$ represent models for outcome confounding bridge functions $h_2(W,M,D,X)$, $h_1(W,D,X)$, and $h_0(W,X)$, respectively, indexed by finite-dimensional parameters $\beta_2$, $\beta_1$, and $\beta_0$. Likewise, $q_0(Z,X;\gamma_0)$, $q_1(Z,D,X;\gamma_1)$, and $q_2(Z,M,D,X;\gamma_2)$ denote models for treatment confounding bridge functions $q_0(Z,X)$, $q_1(Z,D,X)$, and $q_2(Z,M,D,X)$, respectively, indexed by finite-dimensional parameters $\gamma_2$, $\gamma_1$, and $\gamma_0$. It follows from Equations~\eqref{eq_solving h2}-\eqref{eq_solving h0} and \eqref{eq_solving q0 alter}-\eqref{eq_solving q2 alter} that estimates $\hat{\beta}_2$ $\hat{\beta}_1$, $\hat{\beta}_0$, $\hat{\gamma}_0$, $\hat{\gamma}_1$, $\hat{\gamma}_2$ can be obtained as solutions to the estimating equations as follows:
{\small \begin{align}
    \sum_{i=1}^{n} A_i \left [ Y_i - h_2\left ( W_i, M_i, D_i, X_i; \beta_2 \right ) \right ] c_2(Z_i, M_i, D_i, X_i) =& 0, \label{eq: para solve h2} \\ 
    \sum_{i=1}^{n} (1-A_i) \left [ h_2\left ( W_i, M_i, D_i, X_i; \beta_2 \right ) - h_1\left ( W_i, D_i, X_i; \beta_1 \right ) \right ] c_1(Z_i, D_i, X_i) =& 0, \label{eq: para solve h1} \\
    \sum_{i=1}^{n} A_i \left [ h_1\left ( W_i, D_i, X_i; \beta_1 \right ) - h_0\left ( W_i, X_i; \beta_0 \right ) \right ] c_0(Z_i, X_i) =& 0, \label{eq: para solve h0} \\
    \sum_{i=1}^{n} \left [ A_i q_0\left ( Z_i, X_i; \gamma_0 \right ) -1 \right ] b_0(W_i, X_i) =& 0, \label{eq: para solve q0} \\
    \sum_{i=1}^{n} \left [ (1-A_i) q_1\left ( Z_i, D_i, X_i; \gamma_1 \right ) - A_i q_0\left ( Z_i, X_i; \gamma_0 \right ) \right ] b_1(W_i, D_i, X_i) =& 0, \label{eq: para solve q1} \\
    \sum_{i=1}^{n} \left [ A_i q_2\left ( Z_i, M_i, D_i, X_i; \gamma_2 \right ) - (1-A_i) q_1\left ( Z_i, D_i, X_i; \gamma_1 \right ) \right ] b_2(W_i, M_i, D_i, X_i) =& 0, \label{eq: para solve q2}
\end{align}}
where \eqref{eq: para solve h2}-\eqref{eq: para solve h0} are solved sequentially for $(\beta_2, \beta_1, \beta_0)$, and similarly \eqref{eq: para solve q0}-\eqref{eq: para solve q2} are solved sequentially for $(\gamma_0, \gamma_1, \gamma_2)$. The functions $c_2(Z,M,D,X)$, $c_1(Z,D,X)$, $c_0(Z,X)$, $b_0(W,X)$, $b_1(W,D,X)$, and $b_2(W,M,D,X)$ are of the same dimension as their respective parameters $\beta_2$, $\beta_1$, $\beta_0$, $\gamma_0$, $\gamma_1$, and $\gamma_2$. Importantly, integral equations~\eqref{eq_solving q0}-\eqref{eq_solving q2} seem to suggest that the estimation of $\gamma_0$, $\gamma_1$, and $\gamma_2$ would require the postulation of models for $\frac{1}{f(A=1|W, X)}$, $\frac{f(A=1|D, W, X)}{f(A=0|D, W, X)}$, and $\frac{f(A=0|M, D, W, X)}{f(A=1|M, D, W, X)}$, whereas Proposition~\ref{propo-solving q0 q1 q2} indicates that this is not the case by relying on estimating equations~\eqref{eq: para solve q0}-\eqref{eq: para solve q2}. Although locally efficient choices of $c_2$, $c_1$, $c_0$, $b_0$, $b_1$, and $b_2$ are implied in the Appendix of \cite{cui2024semiparametric}, the efficiency gains relative to simple linear choices $c_2(Z, M, D, X) = (1, Z, M, D, X)^T$, $c_1(Z, D, X) = (1, Z, D, X)^T$, $c_0(Z, X) = (1, Z, X)^T$, $b_0(W, X) = (1, W, X)^T$, $b_1(W, D, X) = (1, W, D, X)^T$, and $b_2(W, M, D, X) = (1, W, M, D, X)^T$ are typically modest \citep{stephens2014locally}. We therefore do not pursue locally efficient estimation of nuisance parameters further.

Once we have strategies for estimating nuisance parameters $\beta_2$, $\beta_1$, $\beta_0$, $\gamma_0$, $\gamma_1$, and $\gamma_2$ indexing the confounding bridge functions, one can leverage the identification formulas~\eqref{eq_id h0}, \eqref{eq_id q2}, \eqref{eq_id h1q0}, and \eqref{eq_id h2q1} to construct a proximal outcome regression (P-OR), a proximal inverse probability weighting (P-IPW), and two proximal hybrid (P-hybrid$_1$ and P-hybrid$_2$) estimators of $\psi$, respectively.
{\small
\begin{align*}
    \hat{\psi}_{P-OR} =& \frac{1}{n} \sum_{i=1}^{n} h_0(W_i,X_i; \hat{\beta}_0), \\
    \hat{\psi}_{P-IPW} =& \frac{1}{n} \sum_{i=1}^{n} \left [ A_i Y_i q_2(Z_i,M_i,D_i,X_i; \hat{\gamma}_2) \right ], \\
    \hat{\psi}_{P-hybrid_1} =& \frac{1}{n} \sum_{i=1}^{n} \left [ A_i h_1(W_i,D_i,X_i; \hat{\beta}_1) q_0(Z_i,X_i; \hat{\gamma}_0) \right ], \\
    \hat{\psi}_{P-hybrid_2} =& \frac{1}{n} \sum_{i=1}^{n} \left [ (1-A_i) h_2(W_i,M_i,D_i,X_i; \hat{\beta}_2) q_1(Z_i,D_i,X_i; \hat{\gamma}_1) \right ].
\end{align*}}
Specifically, $\hat{\psi}_{P-OR}$ is a consistent and asymptotically normal (CAN) estimator under the semiparametric model $\cM_1$, $\hat{\psi}_{P-IPW}$ is CAN under model $\cM_2$, $\hat{\psi}_{P-hybrid_1}$ is CAN under model $\cM_3$, and $\hat{\psi}_{P-hybrid_2}$ is CAN under model $\cM_4$. Furthermore, the following assumption yields a proximal quadruply robust (P-quadR) estimator of $\psi$, thereby relaxing parametric modeling assumptions and attaining the semiparametric efficiency bound locally.

\begin{theorem}
    Under typical regularity conditions,
    \begin{align*}
        \hat{\psi}_{P-quadR} = \frac{1}{n}\sum_{i=1}^{n} &\left \{ A_i q_0(Z_i,X_i;\hat{\gamma}_0) \left [ h_1(W_i,D_i,X_i;\hat{\beta}_1) - h_0(W_i,X_i;\hat{\beta}_0) \right ] \right. \\
        &+ (1-A_i) q_1(Z_i,D_i,X_i;\hat{\gamma}_1) \left [ h_2(W_i,M_i,D_i,X_i;\hat{\beta}_2) - h_1(W_i,D_i,X_i;\hat{\beta}_1) \right ] \\
        &+ \left. A_i q_2(Z_i,M_i,D_i,X_i;\hat{\gamma}_2) \left [ Y_i - h_2(W_i,M_i,D_i,X_i;\hat{\beta}_2) \right ] + h_0(W_i,X_i;\hat{\beta}_0) \right\}
    \end{align*}
    is a consistent and asymptotically normal estimator of $\psi$ under the semiparametric union model $\cM_{union} = \cM_{1} \cup \cM_{2} \cup \cM_{3} \cup \cM_{4}$. Additionally, under the semiparametric model $\cM_{sp}$, $\hat{\psi}_{P-quadR}$ is locally efficient at the intersection submodel $\cM_{int} = \cM_{1} \cap \cM_{2} \cap \cM_{3} \cap \cM_{4}$ where Assumption~\ref{asm: regularity} is satisfied.
 \label{thm-robustness}
\end{theorem}

Theorem~\ref{thm-robustness} indicates that valid inference for $\psi$ is possible even when only one of several complex models for confounding bridge functions is correctly specified, and achieves locally efficient estimation when all the models are correctly specified.

\section{Proximal debiased machine learning estimation}\label{sec: nonpara}

We have developed a semiparametric quadruply robust estimator $\hat{\psi}_{P-quadR}$, constructed using the efficient influence function $EIF_{\psi}$, which relies on parametric estimation of nuisance bridge functions. In this section, we extend the developed proximal estimation to a more flexible framework that accommodates nonparametric estimators of bridge functions, seamlessly addressing the obstacle induced by high-dimensional nuisance parameters. A key property of the robust estimation based on efficient influence function is the so-called \textit{Neyman orthogonality} \citep{neyman1979c}, which enables the constructed estimator to remain insensitive to small perturbations in nuisance function estimation. Consequently, noisy estimates of nuisance functions can be used without strongly violating the moment condition of the resulting estimating equation.

We formally propose a proximal debiased machine learning approach in Algorithm~\ref{algo: DML}, combining the double machine learning framework \citep{chernozhukov2018double} with the proximal efficient influence function $EIF_{\psi}$ constructed in Theorem~\ref{thm-EIF}. We employ the cross-fitting technique \citep{schick1986asymptotically, chernozhukov2018double} to mitigate bias due to overfitting and relax regularity conditions required for nuisance function estimation. Algorithm~\ref{algo: DML} integrates modern machine learning estimators of bridge functions $h_2$, $h_1$, $h_0$, $q_2$, $q_1$, and $q_0$, even when these nuisance estimators converge slower than the standard $\sqrt{n}$-rate.

\medskip
\begin{breakablealgorithm}
    \caption{Proximal debiased machine learning for $\psi$}
    \begin{enumerate}
        \item Generate $J$ by duplicating the fold indices $\left \{ 1, 2, ..., L \right \}$ for $\frac{n}{L}$ times, such that $J$ consists of $\frac{n}{L}$ copies of $l$ for any $l \in \left \{ 1, 2, ..., L \right \}$.
        \item Generate $\left \{ J_i \right \}_{i=1}^{n}$ by sampling uniformly from $J$ without replacement, such that $I_l = \left \{ i: J_i = l \right \}$ represents the observation indices belonging to the $l$-th fold, for any $l \in \left \{ 1, 2, ..., L \right \}$ with equal sizes.
        \item \textbf{for} $l= 1, 2, ..., L$, \textbf{do}
        \item \qquad Define $I_l^c := \bigcup_{i\neq l} I_i = \left \{ 1, 2, ..., n \right \} \setminus I_l$, and construct machine learning estimators
        \begin{equation*}
            \hat{\zeta}_l = \hat{\zeta}((\cO_i)_{i\in I_l^c}) = \left( \hat{h}_{2_l}, \hat{h}_{1_l}, \hat{h}_{0_l}, \hat{q}_{2_l}, \hat{q}_{1_l}, \hat{q}_{0_l} \right)
        \end{equation*}
        of confounding bridge functions $\zeta = (h_2, h_1, h_0, q_2, q_1, q_0)$ based on the integral equations~\eqref{eq_solving h2}-\eqref{eq_solving h0} and \eqref{eq_solving q0}-\eqref{eq_solving q2}, which depends only on the subset of data indexed by $I_l^c$.
        \item \qquad Construct the estimator
        \begin{align*}
            \hat{\psi}_{(P-DML)_l} = \frac{L}{n}\sum_{i\in I_l} &\left \{ A_i \hat{q}_{0_l}(Z_i,X_i) \left [ \hat{h}_{1_l}(W_i,D_i,X_i) - \hat{h}_{0_l}(W_i,X_i) \right ] \right. \\
            &+ (1-A_i) \hat{q}_{1_l}(Z_i,D_i,X_i) \left [ \hat{h}_{2_l}(W_i,M_i,D_i,X_i) - \hat{h}_{1_l}(W_i,D_i,X_i) \right ] \\
            &+ \left. A_i \hat{q}_{2_l}(Z_i,M_i,D_i,X_i) \left [ Y_i - \hat{h}_{2_l}(W_i,M_i,D_i,X_i) \right ] + \hat{h}_{0_l}(W_i,X_i) \right\},
        \end{align*}
        where the empirical average depends only on the $l$-th data fold indexed by $I_l$.
        \item \textbf{end for}
        \item Obtain the proximal debiased machine learning estimator by aggregating the above $L$ estimators:
        \begin{equation}
            \hat{\psi}_{P-DML} = \frac{1}{L}\sum_{l=1}^{L} \hat{\psi}_{(P-DML)_l}. \label{eq-DML}
        \end{equation}
    \end{enumerate}
 \label{algo: DML}
\end{breakablealgorithm}
\medskip

Recall that the nuisance bridge functions $h_2$, $h_1$, $h_0$, $q_2$, $q_1$, and $q_0$ are solutions to integral equations and cannot be solved by standard regression methods. Notably, \cite{kallus2021causal} and \cite{ghassami2022minimax} introduced a kernel-based estimation approach for semiparametric proximal causal inference, leveraging nonparametric adversarial learning techniques for solving conditional moment equations \citep{dikkala2020minimax}. Here, we extend this methodology to proximal path-specific inference, presenting a regularized optimization-based approach for estimating bridge functions that satisfy integral equations~\eqref{eq_solving h2}-\eqref{eq_solving h0} and \eqref{eq_solving q0 alter}-\eqref{eq_solving q2 alter}:
{\footnotesize \begin{align*}
    \hat{h}_2 =& \arg\min_{h_2 \in \cH} \sup_{f \in \cF} \E_{n_1} \left \{ \left [ Y - h_2(W,M,D,X) \right ] f(Z,M,D,X) - f^2(Z,M,D,X) \right \} - \lambda_{\cF}^{h_2} \left \| f \right \|_{\cF}^2 + \lambda_{\cH}^{h_2} \left \| h_2 \right \|_{\cH}^2, \\
    \hat{h}_1 =& \arg\min_{h_1 \in \cH} \sup_{f \in \cF} \E_{n_0} \left \{ \left [ \hat{h}_2(W,M,D,X) - h_1(W,D,X) \right ] f(Z,D,X) - f^2(Z,D,X) \right \} - \lambda_{\cF}^{h_1} \left \| f \right \|_{\cF}^2 + \lambda_{\cH}^{h_1} \left \| h_1 \right \|_{\cH}^2, \\
    \hat{h}_0 =& \arg\min_{h_0 \in \cH} \sup_{f \in \cF} \E_{n_1} \left \{ \left [ \hat{h}_1(W,D,X) - h_0(W,X) \right ] f(Z,X) - f^2(Z,X) \right \} - \lambda_{\cF}^{h_0} \left \| f \right \|_{\cF}^2 + \lambda_{\cH}^{h_0} \left \| h_0 \right \|_{\cH}^2, \\
    \hat{q}_0 =& \arg\min_{q_0 \in \cQ} \sup_{f \in \cF} 
    \E_n \left \{ \left [ A q_0(Z,X) -1 \right ] f(W,X) - f^2(W,X) \right \} - \lambda_{\cF}^{q_0} \left \| f \right \|_{\cF}^2 + \lambda_{\cQ}^{q_0} \left \| q_0 \right \|_{\cQ}^2, \\
    \hat{q}_1 =& \arg\min_{q_1 \in \cQ} \sup_{f \in \cF} 
    \E_n \left \{ \left [ (1-A) q_1(Z,D,X) - A \hat{q}_0(Z,X) \right ] f(W,D,X) - f^2(W,D,X) \right \} - \lambda_{\cF}^{q_1} \left \| f \right \|_{\cF}^2 + \lambda_{\cQ}^{q_1} \left \| q_1 \right \|_{\cQ}^2, \\
    \hat{q}_2 =& \arg\min_{q_2 \in \cQ} \sup_{f \in \cF} \E_n \left \{ \left [ A q_2(Z,M,D,X) - (1-A) \hat{q}_1(Z,D,X) \right ] f(W,M,D,X) - f^2(W,M,D,X) \right \} - \lambda_{\cF}^{q_2} \left \| f \right \|_{\cF}^2 + \lambda_{\cQ}^{q_2} \left \| q_2 \right \|_{\cQ}^2,
\end{align*}}
where $\cH$, $\cQ$, and $\cF$ represent normed function spaces with norms $\left \| . \right \|_{\cH}$, $\left \| . \right \|_{\cQ}$, and $\left \| . \right \|_{\cF}$, and $\E_n$, $\E_{n_1}$, and $\E_{n_0}$ denote empirical expectations over the whole data and the subsets of data with $A=1$ and $A=0$, respectively. A detailed analysis regarding the convergence rates of the proposed minimax estimators for nuisance functions has been discussed in the literature \citep{dikkala2020minimax, kallus2021causal, ghassami2022minimax}.

\begin{definition}
    In the cross-fitting procedure, $n^{-\tau_{h_2}}$, $n^{-\tau_{h_1}}$, $n^{-\tau_{h_0}}$, $n^{-\tau_{q_2}}$, $n^{-\tau_{q_1}}$, and $n^{-\tau_{q_0}}$, with positive constants $\tau_{h_2}, \tau_{h_1}, \tau_{h_0}, \tau_{q_2}, \tau_{q_1}, \tau_{q_0} >0$, are said to be the convergence rates of $\hat{h}_2$ to $h_2$, $\hat{h}_1$ to $h_1$, $\hat{h}_0$ to $h_0$, $\hat{q}_2$ to $q_2$, $\hat{q}_1$ to $q_1$, and $\hat{q}_0$ to $q_0$ in the root mean square error sense, respectively, if, for every data fold $\left \{ I_l \right \}_{l=1}^{L}$ and $n \rightarrow \infty$,
    {\small \begin{align}
        \sqrt{\E\left [ \left | \hat{h}_{2_l}(W,M,D,X) - h_2(W,M,D,X) \right |^2 \right ]} =& o\left ( n^{-\tau_{h_2}} \right ), \label{limit: h2} \\
        \sqrt{\E\left [ \left | \hat{h}_{1_l}(W,D,X) - h_1(W,D,X) \right |^2 \right ]} =& o\left ( n^{-\tau_{h_1}} \right ), \label{limit: h1} \\
        \sqrt{\E\left [ \left | \hat{h}_{0_l}(W,X) - h_0(W,X) \right |^2 \right ]} =& o\left ( n^{-\tau_{h_0}} \right ), \label{limit: h0} \\
        \sqrt{\E\left [ \left | \hat{q}_{0_l}(Z,X) - q_0(Z,X) \right |^2 \right ]} =& o\left ( n^{-\tau_{q_0}} \right ), \label{limit: q0} \\
        \sqrt{\E\left [ \left | \hat{q}_{1_l}(Z,D,X) - q_1(Z,D,X) \right |^2 \right ]} =& o\left ( n^{-\tau_{q_1}} \right ), \label{limit: q1} \\
        \sqrt{\E\left [ \left | \hat{q}_{2_l}(Z,M,D,X) - q_2(Z,M,D,X) \right |^2 \right ]} =& o\left ( n^{-\tau_{q_2}} \right ). \label{limit: q2}
    \end{align}}
 \label{def: RMSE}
\end{definition}

The following theorem establishes that the proximal debiased machine learning estimator $\hat{\psi}_{P-DML}$ constructed in Algorithm~\ref{algo: DML} remains $\sqrt{n}$-consistent and asymptotically normal under mild regularity conditions, even if nuisance functions are estimated using machine learning methods that do not converge at the $\sqrt{n}$-rate. Before stating the theorem, we impose a regularity assumption to ensure that the conditional mean-zero terms in \eqref{eq_solving h2}-\eqref{eq_solving h0} and \eqref{eq_solving q0 alter}-\eqref{eq_solving q2 alter} for defining nuisance functions have finite conditional second moments.

\begin{assumption}(Finite second-order moments): For some constant $\eta >0$,
    {\small \begin{align*}
        \E\left [ \left ( Y - h_2(W, M, D, X) \right )^2 \big| A=1, M, D, Z, X \right ] \leq& \eta, \\
        \E\left [ \left ( h_2(W, M, D, X) - h_1(W, D, X) \right )^2 \big| A=0, D, Z, X \right ] \leq& \eta, \\
        \E\left [ \left ( h_1(W, D, X) - h_0(W, X) \right )^2 \big| A=1, Z, X \right ] \leq& \eta, \\
        \E\left[ \left( A q_0(Z,X) -1 \right)^2 \big|W, X \right] \leq& \eta, \\
        \E\left [ \left( (1-A) q_1(Z,M,X) - A q_0(Z,X) \right)^2 \big|W,D,X \right ] \leq& \eta, \\
        \E\left [ \left( A q_2(Z,M,D,X) - (1-A) q_1(Z,M,X) \right)^2 \big|W,M,D,X \right ] \leq& \eta.
    \end{align*}}
  \label{asm: finite second-order moments}
\end{assumption}

Note that the convergence rates of machine learning estimators for nuisance functions are not necessarily required to attain the $\sqrt{n}$-rate in the root mean square error sense, that is, $\max\left \{ \tau_{h_2}, \tau_{h_1}, \tau_{h_0}, \tau_{q_2}, \tau_{q_1}, \tau_{q_0} \right \} < 1/2$. Assumption~\ref{asm: convergence rates} enables the sums of some combinations of these convergence rates to possess a lower bound.

\begin{assumption}(Convergence rates of nuisance functions):
    \begin{equation}
        \min\left \{ \tau_{h_2} + \tau_{q_1}, \tau_{h_2} + \tau_{q_2}, \tau_{h_1} + \tau_{q_0}, \tau_{h_1} + \tau_{q_1}, \tau_{h_0} + \tau_{q_0} \right \} \geq 1/2.
    \end{equation}
  \label{asm: convergence rates}
\end{assumption}

\vspace{-0.5cm}
We are now in a position to establish the theorem, guaranteeing that the proximal debiased machine learning Algorithm~\ref{algo: DML} achieves the $\sqrt{n}$-consistency and asymptotic normality, even when machine learning estimation of nuisance functions does not attain the parametric $\sqrt{n}$-convergence.

\begin{theorem}
    Suppose that $\hat{h}_2$, $\hat{h}_1$, $\hat{h}_0$, $\hat{q}_2$, $\hat{q}_1$, and $\hat{q}_0$ are machine learning estimators of confounding bridge functions $h_2$, $h_1$, $h_0$, $q_2$, $q_1$, and $q_0$ with convergence rates $n^{-\tau_{h_2}}$, $n^{-\tau_{h_1}}$, $n^{-\tau_{h_0}}$, $n^{-\tau_{q_2}}$, $n^{-\tau_{q_1}}$, and $n^{-\tau_{q_0}}$ in the root mean square error sense, satisfying \eqref{limit: h2}-\eqref{limit: q2}, respectively. Under Assumption~\ref{asm: finite second-order moments}, the estimator $\hat{\psi}_{P-DML}$ given by \eqref{eq-DML} constructed in Algorithm~\ref{algo: DML} satisfies the following asymptotic equality:
    \begin{equation*}
        \sqrt{n}\left ( \hat{\psi}_{P-DML} - \psi \right ) = \frac{1}{\sqrt{n}}\sum_{i=1}^{n} EIF_{\psi}(\cO_i) + o_p\left ( n^{-\alpha} \right ), 
    \end{equation*}
    where {\small $$\alpha = \min\left \{ \tau_{h_2}, \tau_{h_1}, \tau_{h_0}, \tau_{q_2}, \tau_{q_1}, \tau_{q_0}, \tau_{h_2}+\tau_{q_2}-\frac{1}{2}, \tau_{h_2}+\tau_{q_1}-\frac{1}{2}, \tau_{h_1}+\tau_{q_0}-\frac{1}{2}, \tau_{h_1}+\tau_{q_1}-\frac{1}{2}, \tau_{h_0}+\tau_{q_0}-\frac{1}{2} \right \}.$$}
    Furthermore, if the convergence rates satisfy Assumption~\ref{asm: convergence rates}, $\hat{\psi}_{P-DML}$ would be a $\sqrt{n}$-consistent and asymptotically normal estimator of $\psi$, satisfying
    \begin{equation*}
        \sqrt{n}\left ( \hat{\psi}_{P-DML} - \psi \right ) = \frac{1}{\sqrt{n}}\sum_{i=1}^{n} EIF_{\psi}(\cO_i) + o_p\left ( 1 \right ),
    \end{equation*}
    and $\sqrt{n}\left ( \hat{\psi}_{P-DML} - \psi \right )$ converges in distribution to the normal distribution $\cN\left ( 0, \E\left [ EIF^2_{\psi}(\cO) \right ] \right )$, where $EIF_{\psi}(\cO)$ is the efficient influence function characterized in Theorem~\ref{thm-EIF}.
  \label{thm-DML}
\end{theorem}

A notable special case in which Assumption~\ref{asm: convergence rates} holds is when $\tau_{h_2} = \tau_{h_1} = \tau_{h_0} = \tau_{q_2} = \tau_{q_1} = \tau_{q_0} = 1/4$, implying that the nuisance estimators $\hat{h}_2$, $\hat{h}_1$, $\hat{h}_0$, $\hat{q}_2$, $\hat{q}_1$, and $\hat{q}_0$ are only required to achieve an $o\left(n^{-1/4}\right)$ convergence rate in the root mean square error sense. In contrast, parametric models for nuisance parameters typically attain an $O\left ( 1/\sqrt{n} \right )$ convergence rate. Therefore, Theorem~\ref{thm-DML} accommodates scenarios where $\hat{h}_2$, $\hat{h}_1$, $\hat{h}_0$, $\hat{q}_2$, $\hat{q}_1$, and $\hat{q}_0$ converge at a rate significantly slower than the parametric benchmark. The $o\left(n^{-1/4}\right)$ convergence rate is achievable for various machine learning methods \citep{chen1999improved, Belloni2013least} under appropriate structural assumptions on nuisance parameters.

\section{Numerical experiments}

We conduct two simulation studies to evaluate the performance of the proposed estimators\footnote{The codes are available at \href{https://github.com/YangBAI330/Proximal-Path-Specific-Inference.git}{GitHub}.}. In Section~\ref{sec: simu semiparametric}, we employ parametric models to estimate the nuisance bridge functions, to illustrate the robustness of the semiparametric proximal estimators constructed in Section~\ref{sec: semipara} under varying degrees of model misspecification. In Section~\ref{sec: simu nonparametric}, we use the nonparametric regularized minimax approach to estimate the nuisance bridge functions, to evaluate the finite-sample performance of the proximal debiased machine learning estimators designed in Section~\ref{sec: nonpara}.

\subsection{Simulation I: semiparametric estimation}\label{sec: simu semiparametric}

In this section, we conduct a simulation study to evaluate the finite-sample performance of our new proposed estimators for path-specific effects under varying degrees of model misspecification. The data-generating mechanism of $(Y, A, D, M, W, Z, X, U)$ follows Assumptions~\ref{asm: consistency}-\ref{asm: completeness for W} for the proximal causal path-specific framework described in Section~\ref{sec: identification}. The specific data-generating process is provided in the supplementary material.

We evaluate the performance of proposed semiparametric proximal estimators $\hat{\psi}_{P-OR}$, $\hat{\psi}_{P-IPW}$, $\hat{\psi}_{P-hybrid_1}$, $\hat{\psi}_{P-hybrid_2}$, and $\hat{\psi}_{P-quadR}$. Confounding bridge functions $\left\{ h_2, h_1, h_0 \right\}$ and $\left\{ q_0, q_1, q_2 \right\}$ are obtained by sequentially solving the corresponding estimating equations~\eqref{eq: para solve h2}-\eqref{eq: para solve h0} and \eqref{eq: para solve q0}-\eqref{eq: para solve q2}, respectively, to yield their estimates under specified models. The model specifications of bridge functions are provided and justified in the supplementary material. We evaluate these estimators in the following five scenarios, where models $\cM_1$, $\cM_2$, $\cM_3$, and $\cM_4$ are misspecified to varying degrees:
\begin{enumerate}
    \item $\cM_1 \cap \cM_2 \cap \cM_3 \cap \cM_4$: all the confounding bridge functions are correctly modeled;
    \item $\cM_1$ correct: $q_0$, $q_1$, and $q_2$ are misspecified;
    \item $\cM_2$ correct: $h_2$, $h_1$, and $h_0$ are misspecified;
    \item $\cM_3$ correct: $h_0$, $q_1$, and $q_2$ are misspecified;
    \item $\cM_4$ correct: $h_1$, $h_0$, and $q_2$ are misspecified.
\end{enumerate}
The simulations were conducted with a sample size $n = 1000$, repeated $500$ times. The confidence intervals were computed using the nonparametric bootstrap method.

\begin{table}[p]
	\centering
    \resizebox{0.7\textwidth}{!}{
		\begin{tabular}{ccccccccccc}
			\toprule
			Scenarios & \multicolumn{4}{c}{Models} &  & \multicolumn{5}{c}{Estimators} \\
			\cmidrule{2-5}
			\cmidrule{7-11}
			&\textbf{$\mathcal{M}_1$} & \textbf{$\mathcal{M}_2$} & \textbf{$\mathcal{M}_3$} & \textbf{$\mathcal{M}_4$} & & \textit{P-OR} & \textit{P-IPW} & \textit{P-hybrid$_1$} & \textit{P-hybrid$_2$} & \textit{P-quadR} \\
			\cmidrule{2-5}
			\cmidrule{7-11}
			\multicolumn{11}{c}{ \textbf{Bias} } \\
			1 & \checkmark & \checkmark & \checkmark & \checkmark & & $\mathbf{0.002}$ & $\mathbf{0.004}$ & $\mathbf{0.000}$ & $\mathbf{-0.001}$ & $\mathbf{0.003}$ \\
			2 & \checkmark & $\times$ & $\times$ & $\times$ & & $\mathbf{-0.010}$ & $-3.089$ & $-0.302$ & $-0.136$ & $\mathbf{-0.013}$ \\
			3 & $\times$ & \checkmark & $\times$ & $\times$ & & $-0.334$ & $\mathbf{-0.011}$ & $-0.191$ & $-0.210$ & $\mathbf{-0.009}$ \\
			4 & $\times$ & $\times$ & \checkmark & $\times$ & & $-0.419$ &$-3.056$ & $\mathbf{-0.005}$ & $0.147$ & $\mathbf{-0.014}$ \\
			5 & $\times$ & $\times$ & $\times$ & \checkmark & & $-0.185$ & $-2.892$ & $0.005$ & $\mathbf{-0.015}$ & $\mathbf{-0.008}$ \\
					
			\\
			\multicolumn{11}{c}{ \textbf{MSE} } \\
			1 & \checkmark & \checkmark & \checkmark & \checkmark & & $\pmb{0.031}$ & $\pmb{0.047}$ & $\pmb{0.032}$ & $\pmb{0.033}$ & $\pmb{0.043}$ \\
			2 & \checkmark & $\times$ & $\times$ & $\times$ & & \pmb{$0.032$} & $10.097$ & $0.128$ & $0.085$ & \pmb{$0.036$} \\
			3 & $\times$ & \checkmark & $\times$ & $\times$ & & $0.153$ & $\pmb{0.047}$ & $0.077$ & $0.086$ & $\pmb{0.047}$ \\
			4 & $\times$ & $\times$ & \checkmark & $\times$ & & $0.228$ & $10.031$ & $\pmb{0.035}$ & $0.124$ & $\pmb{0.047}$ \\
			5 & $\times$ & $\times$ & $\times$ & \checkmark & & $0.087$ & $9.454$ & $0.049$ & $\pmb{0.043}$ & $\pmb{0.044}$ \\

            \\
			\multicolumn{11}{c}{ \textbf{Coverage} } \\
			1 & \checkmark & \checkmark & \checkmark & \checkmark & & \pmb{$95.0\%$} & \pmb{$95.8\%$} & \pmb{$95.2\%$} & \pmb{$95.6\%$} & \pmb{$96.2\%$} \\
			2 & \checkmark & $\times$ & $\times$ & $\times$ & & \pmb{$95.2\%$} & $5.8\%$ & $64.8\%$ & $84.2\%$ & \pmb{$95.4\%$} \\
			3 & $\times$ & \checkmark & $\times$ & $\times$ & & $59.6\%$ & \pmb{$96.0\%$} & $79.4\%$ & $79.0\%$ & \pmb{$95.4\%$} \\
			4 & $\times$ & $\times$ & \checkmark & $\times$ & & $45.4\%$ & $6.0\%$ & \pmb{$96.0\%$} & $84.6\%$ & \pmb{$95.2\%$} \\
			5 & $\times$ & $\times$ & $\times$ & \checkmark & & $82.2\%$ & $11.6\%$ & $87.2\%$ & \pmb{$95.2\%$} & \pmb{$96.6\%$} \\
            
            \\
			\multicolumn{11}{c}{ \textbf{Length} } \\
			1 & \checkmark & \checkmark & \checkmark & \checkmark & & \pmb{$0.705$} & \pmb{$0.901$} & \pmb{$0.712$} & \pmb{$0.755$} & \pmb{$0.854$} \\
			2 & \checkmark & $\times$ & $\times$ & $\times$ & & \pmb{$0.699$} & $1.944$ & $0.752$ & $1.587$ & \pmb{$0.790$} \\
			3 & $\times$ & \checkmark & $\times$ & $\times$ & & $0.818$ & \pmb{$0.840$} & $0.806$ & $0.818$ & \pmb{$0.839$} \\
			4 & $\times$ & $\times$ & \checkmark & $\times$ & & $0.866$ & $2.518$ & \pmb{$0.708$} & $1.868$ & \pmb{$0.892$} \\
			5 & $\times$ & $\times$ & $\times$ & \checkmark & & $0.909$ & $2.310$ & $0.870$ & \pmb{$0.766$} &  \pmb{$0.799$} \\
			\bottomrule
			\end{tabular}}
		\caption{Simulation results across Scenarios 1-5. Bias: Monte Carlo bias; MSE: mean squared error; Coverage: $95\%$ confidence interval coverage rate; Length: average $95\%$ confidence interval length. Emboldened values come from those consistent estimators in the corresponding scenarios.}
    \label{table: simulation}
\end{table}

The simulation results are summarized in Table~\ref{table: simulation}. As anticipated, $\hat{\psi}_{P-OR}$ exhibits small bias and achieves nominal confidence interval coverage only in Scenarios 1 and 2, where $h_2$, $h_1$, and $h_0$ are correctly specified. In contrast, $\hat{\psi}_{P-IPW}$ shows small bias and nominal confidence interval coverage only in Scenarios 1 and 3, provided that $q_0$, $q_1$, and $q_2$ are correctly specified. Likewise, $\hat{\psi}_{P-hybrid_1}$ achieves small bias and nominal confidence interval coverage only in Scenarios 1 and 4 when $h_2$, $h_1$, and $q_0$ are correctly specified, while $\hat{\psi}_{P-hybrid_2}$ attains small bias and nominal confidence interval coverage only in Scenarios 1 and 5 when $h_2$, $q_0$, and $q_1$ are correctly specified. Notably, consistent with Theorem~\ref{thm-robustness}, the quadruply robust estimator $\hat{\psi}_{P-quadR}$ maintains low bias and produces nominal confidence interval coverage across all scenarios. This robustness underscores the utility of $\hat{\psi}_{P-quadR}$ in accommodating varying model misspecifications and maintaining valid inference in complex settings.

\subsection{Simulation II: nonparametric estimation}\label{sec: simu nonparametric}

In this section, we evaluate the performance of the proximal debiased machine learning Algorithm~\ref{algo: DML} in the cross-fitting procedure. Specifically, we use the reproducing kernel Hilbert space (RKHS) as the function space for the nonparametric regularized minimax learning estimation of nuisance bridge functions. The distribution of $(Y, A, D, M, W, Z, X, U)$ follows Assumptions~\ref{asm: consistency}-\ref{asm: completeness for W} for the proximal causal path-specific framework described in Section~\ref{sec: identification}. The details of the data-generating process are presented in the supplementary material. To evaluate finite-sample performances of proposed estimators, we consider three cases with sample sizes $n \in \left \{ 200, 1000, 2000 \right \}$, respectively, and repeat each experiment for $300$ times.

In contrast to the specification of parametric models for nuisance bridge functions in Section~\ref{sec: simu semiparametric}, we instead employ the regularized optimization-based approach developed in Section~\ref{sec: nonpara} to obtain nonparametric estimators of nuisance functions, with RKHS as normed spaces for nuisance functions $h_2$, $h_1$, $h_0$, $q_2$, $q_1$, and $q_0$. The hyperparameters of the estimation approach are selected via cross-validation. Along with the proximal debiased machine learning estimator $\hat{\psi}_{P-DML}$ based on the efficient influence function, we also demonstrate the finite-sample performances of other proximal strategies (P-OR, P-IPW, P-hybrid$_1$, and P-hybrid$_2$) using nonparametric estimators of nuisance functions in the cross-fitting procedure. To distinguish them from their semiparametric counterparts illustrated by Section~\ref{sec: simu semiparametric}, with a slight abuse of notation, in this section, we rename the four strategies (por, pipw, phybrid$_1$, and phybrid$_2$) to emphasize that they are constructed in the context of Algorithm~\ref{algo: DML}. 

\begin{figure}[ht]
    \centering
    \includegraphics[width=0.8\textwidth]{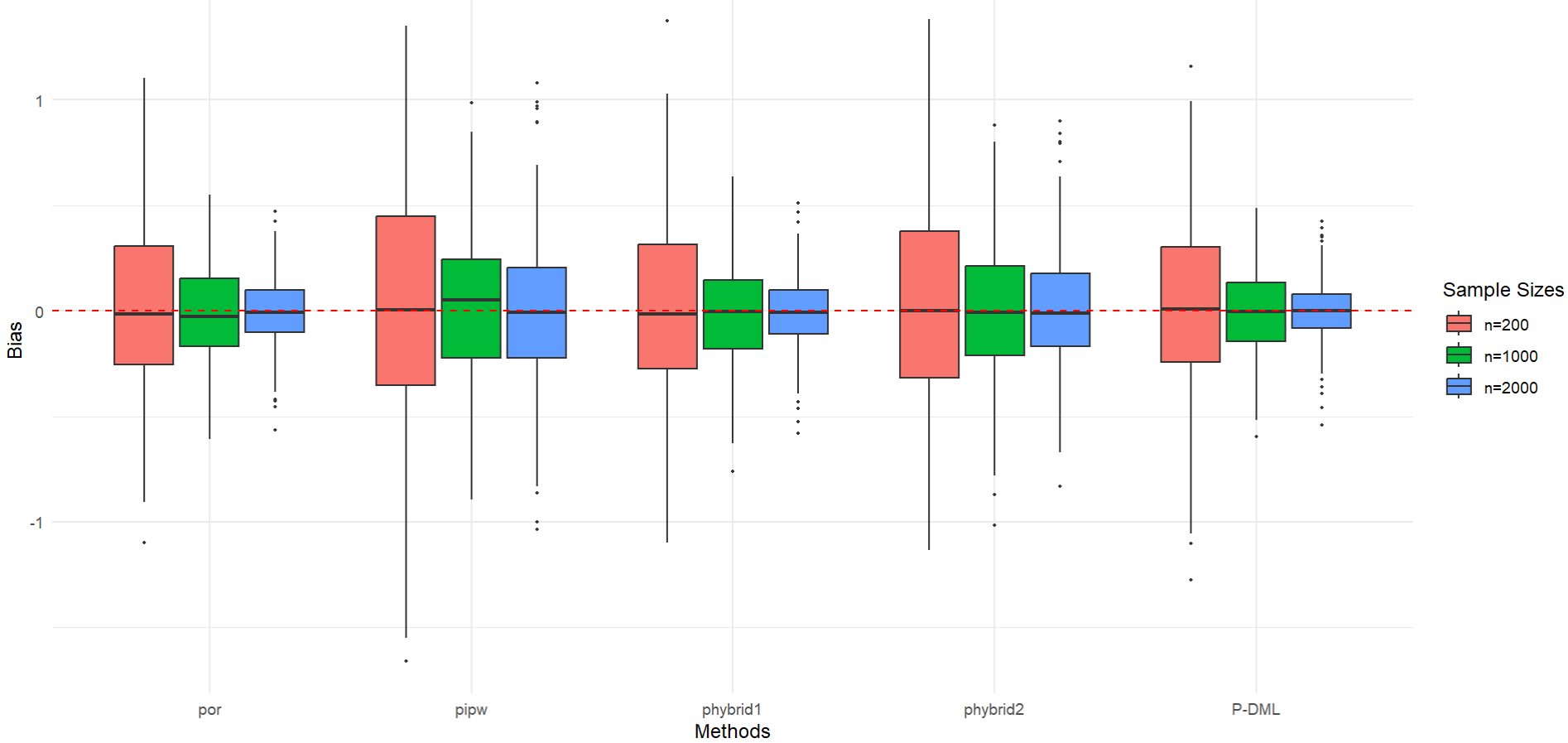}
    \caption{Comparison of bias across sample sizes ($n=200$, $n=1000$, and $n=2000$) for estimation methods (por, pipw, phybrid$_1$, phybrid$_2$, and P-DML). The dashed red line indicates zero bias.}
    \label{fig: simu_bias_boxplot}
\end{figure}

Figure~\ref{fig: simu_bias_boxplot} presents the boxplots of bias results for the proposed nonparametric proximal estimators across different sample sizes. Each box is generated from $300$ experiments. We compare the performance of the proximal debiased machine learning estimator (P-DML) with por, pipw, phybrid$_1$, and phybrid$_2$ estimators. Most methods appear to be centered around the dashed red line (bias = 0). As anticipated, P-DML outperforms the other estimators in terms of the distribution of bias for every finite-sample case, with median biases near zero even at the smallest sample size ($n=200$). Figure~\ref{fig: simu_bias_boxplot} also demonstrates the effect of sample size on the performance of nonparametric proximal estimators. As the sample size $n$ increases, the bias of each estimator decreases. While all methods converge toward zero bias by sample size, the P-DML estimator exhibits the narrowest boxes, suggesting it may be the most statistically efficient.

\begin{table}[ht]
    \centering
    \setlength{\tabcolsep}{8pt} 
    \resizebox{0.8\textwidth}{!}{
    \begin{tabular}{l cc cc cc cc cc}
        \toprule
        \multirow{2}{*}{sample size ($n$)} & \multicolumn{2}{c}{\textit{por}} & \multicolumn{2}{c}{\textit{pipw}} & \multicolumn{2}{c}{\textit{phybrid$_1$}} & \multicolumn{2}{c}{\textit{phybrid$_2$}} & \multicolumn{2}{c}{\textit{pDML}} \\
        \cmidrule(lr){2-3} \cmidrule(lr){4-5} \cmidrule(lr){6-7} \cmidrule(lr){8-9} \cmidrule(lr){10-11}
        & Bias & MSE & Bias & MSE & Bias & MSE & Bias & MSE & Bias & MSE \\
        \midrule
        200  & 0.009  & 0.139 & 0.032  & 0.341 & 0.011  & 0.165 & 0.021  & 0.238 & 0.011  & 0.141 \\ 
        1000 & -0.013 & 0.049 & 0.016  & 0.129 & -0.011 & 0.054 & -0.008 & 0.100 & -0.008 & 0.038 \\
        2000 & -0.004 & 0.027 & -0.002 & 0.121 & -0.004 & 0.029 & 0.005  & 0.079 & -0.003 & 0.020 \\
        \bottomrule
    \end{tabular}}
    \caption{Results from the debiased machine learning Algorithm~\ref{algo: DML}, combined with the regularized minimax-based estimated bridge functions. Bias: Monte Carlo bias; MSE: mean squared error.}
    \label{table: DML bias mse}
\end{table}

Table~\ref{table: DML bias mse} summarizes the average performances of the proposed proximal estimators over the $300$ experiments under varying sample sizes, including the Monte Carlo bias and mean square error (MSE). Specifically, the P-DML estimator consistently yields the lowest MSE and near-lowest Monte Carlo bias across all sample sizes, indicating superior statistical efficiency compared to other estimators. Additionally, all estimators exhibit consistent reduction in Monte Carlo bias and MSE as $n$ increases, and most methods exhibit low bias even in small samples ($n = 200$). This confirms the consistency and robust performance of the nonparametric proximal estimators.

\section{Real data analysis}

We apply the proposed proximal estimators of the path-specific effect in the latest CDC WONDER Natality data\footnote{The dataset is public (\href{https://wonder.cdc.gov/natality.html}{\textit{https://wonder.cdc.gov/natality.html}}).} for 2016-2024, which is a national set of U.S. birth records compiled from birth certificates of U.S. residents. We focus our analysis on the subset of participants who are Asian and do not reside in California, since California no longer provides record-level data on the mother's marital status for births occurring in California beginning in 2017 due to state statutory restrictions. Subjects with unknown or not stated data ($<1.8\%$ of the samples) are excluded, and the total sample size is $174,042$.

We aim to investigate the path-specific effect of prenatal care ($A$) on preterm birth ($Y$), mediated by preeclampsia ($M$) but not through smoking status during pregnancy ($D$). As pointed out by \cite{xia2023identification}, the smoking status during pregnancy is a potential treatment-induced confounder in the relationship between preeclampsia and preterm birth. The measurement of adequate prenatal care ($A=1$) follows the Adequacy of Prenatal Care Utilization (APNCU) index \citep{kotelchuck1994evaluation}, which is determined by the month prenatal care began, the number of prenatal visits, and the gestational age at the time of delivery. Preterm birth is determined using the Obstetric Estimate (OE) \citep{martin2015measuring} of gestational age, where we use $Y=0$ to stand for suffering from preterm birth (versus $Y=1$ denoting not preterm birth). The observed covariates ($X$) include maternal education level and prior preterm birth. Section~\ref{sec: introduction} briefly summarizes the current study progress and limitations in existing literature \citep{vansteelandt2012natural, vanderweele2014effect, xia2023identification}. The detailed discussion is provided in the supplementary material. Accordingly, the unmeasured confounders that are not collected in the CDC WONDER Natality dataset (such as household income, housing stability, chronic stress, exposure to violence, discrimination, racism, and adverse childhood experiences) unavoidably plague the relationships among $A$, $D$, $M$, and $Y$, which are beyond the scope of \cite{miles2017quantifying, miles2020semiparametric} that only account for the unmeasured confounding restricted in the $D-Y$ relationship.

Given the inevitability of unmeasured confounding in practical applications, a viable strategy involves selecting baseline covariates that can serve as proxy variables for the latent confounding mechanisms. We provide a detailed discussion on the criteria and procedure for proxy variable selection in the Supplementary Material. Accordingly, marital status ($Z$) mainly reflects latent socioeconomic support structures, which predict the adequacy of prenatal care ($A$), and do not have direct causal effects on smoking during pregnancy ($D$), preeclampsia ($M$), and preterm birth ($Y$) outside of its pathway through $A$ and unmeasured confounders. Furthermore, pre-pregnancy hypertension ($W$) reflects latent cardiovascular susceptibility, which has no direct effect on $A$, $D$, and $M$; on the other hand, $W$ is determined prior to pregnancy and thus cannot be affected by these measurements during pregnancy. Although epidemiological studies find associations between $Z$ and $W$, these relationships are best understood as their common unmeasured confounders. 

\begin{table}[ht]
    \centering
    \resizebox{0.8\textwidth}{!}{
    \begin{tabular}{l ccccc}
        \toprule
         & P-OR & P-IPW & P-hybrid$_1$ & P-hybrid$_2$ & P-quadR \\
        \midrule
        \multicolumn{6}{l}{\textbf{Panel A: Decreased Risk ($\mathcal{P}_{AMY}$)}} \\
        Estimate & $0.03\%$ & $0.05\%$ & $0.04\%$ & $0.03\%$ & \pmb{$0.06\%$} \\
        Bootstrap SE & $0.01\%$ & $0.02\%$ & $0.02\%$ & $0.02\%$ & $0.01\%$ \\
        Bootstrap CI & $(0.01\%, 0.07\%)$ & $(0.01\%, 0.07\%)$ & $(0.02\%, 0.09\%)$ & $(-0.00\%, 0.06\%)$ & \pmb{$(0.03\%, 0.07\%)$} \\
        \midrule
        \multicolumn{6}{l}{\textbf{Panel B: Reduction in Odds ($\mathcal{R}_{AMY}$)}} \\
        Estimate & $0.59\%$ & $0.78\%$ & $0.68\%$ & $0.47\%$ & \pmb{$0.98\%$} \\
        Bootstrap SE & $0.24\%$ & $0.28\%$ & $0.26\%$ & $0.28\%$ & $0.18\%$ \\
        Bootstrap CI & $(0.21\%, 1.12\%)$ & $(0.21\%, 1.17\%)$ & $(0.39\%, 1.46\%)$ & $(-0.07\%, 1.01\%)$ & \pmb{$(0.61\%, 1.27\%)$} \\
        \bottomrule
    \end{tabular}}
    \caption{Estimated path-specific effects with standard errors and $95\%$ confidence intervals of adequate prenatal care on preterm birth, mediated by preeclampsia but not through smoking.}
    \label{table: real data}
\end{table}

Table~\ref{table: real data} presents the estimates for $\mathcal{P}_{AMY}$ and the reduction in odds, $\mathcal{R}_{AMY}$, using bootstrap standard errors and $95\%$ confidence intervals for inference. The quadruply robust estimator indicates that the path-specific effect of adequate prenatal care, through preeclampsia excluding the smoking pathway, decreases preterm birth risk by $0.06\%~(0.03\%, 0.07\%)$. To facilitate comparison with existing literature, we also report the reduction in odds of preterm birth:
{\small $$\mathcal{R}_{AMY} = 1- \frac{P\left( Y\left ( 1, D(1), M\left ( 1, D(1) \right ) \right ) =0 \right) /  P\left( Y\left ( 1, D(1), M\left ( 1, D(1) \right ) \right ) =1 \right)}{P\left( Y\left ( 1, D(1), M\left ( 0, D(1) \right ) \right ) =0 \right) /  P\left( Y\left ( 1, D(1), M\left ( 0, D(1) \right ) \right ) =1 \right)}.$$}
Notably, the P-quadR estimator yields the narrowest confidence intervals and the largest estimates for both $\mathcal{P}_{AMY}$ and $\mathcal{R}_{AMY}$. It estimates that the indirect effect of adequate prenatal care, through preeclampsia but not via smoking, amounts to a reduction in the odds of preterm birth with $0.98\%~(0.61\%, 1.27\%)$, which is consistent with the interventional effect results in \cite{vanderweele2014effect}. In contrast to previous analyses \citep{vanderweele2014effect, xia2023identification} that suggested adequate prenatal care through preeclampsia increases preterm birth risk, our results better align with epidemiological intuition. This is achieved by leveraging proxy variables to account for complex unmeasured confounding, isolating the specific path through $M$ from $D$, and relaxing stringent heterogeneity assumptions required by earlier methods \citep{xia2023identification}.

\section*{Supplementary material}
\label{SM}

The Supplementary Material includes proofs of the theorems, details of the data-generating processes in simulation studies, details regarding the bridge function choices in simulation I, and supplements to the real data analysis.

\bibliography{paper-ref}

\end{document}